\documentclass[prb,twocolumn,superscriptaddress]{revtex4-2}
\usepackage{ gensymb }
\usepackage{amsmath}
\usepackage{amssymb}
\usepackage{amsthm}
\usepackage{amsfonts}
\usepackage[version=4]{mhchem}
\usepackage{listings}
\usepackage{enumerate}
\usepackage{latexsym}
\usepackage{psfrag}
\usepackage{bm} 
\usepackage[all]{xy}
\usepackage{subfigure}
\usepackage[colorlinks]{hyperref}
\usepackage{color}
\usepackage{mathtools}
\usepackage{array}
\usepackage{placeins}
\begin{document}
\title{Magnetic phase transitions in the triangular-lattice spin-1 dimer compound \ce{K2Ni2(SeO3)3}}
\author{Lei Yue}
\author{Ziyou Lu}
\affiliation{Department of Physics, Southern University of Science and
  Technology, Shenzhen, 518055, China}
\author{Kun Yan}
\affiliation{School of Science and State Key Laboratory on Tunable Laser
  Technology and Ministry of Industry and Information Technology Key Lab of 
  Micro-Nano Optoelectronic Information System, Harbin Institute of Technology,
  Shenzhen 518055, China }
\author{Le Wang}
\email{wangl36@sustech.edu.cn}
\author{Shu Guo}
\author{Ruixin Guo}
\affiliation{Shenzhen Institute for Quantum Science and Engineering, Southern University of Science and Technology, Shenzhen 518055, China}
\affiliation{International Quantum Academy, Shenzhen 518048, China}
\author{Peng Chen}
\affiliation{Songshan Lake Materials Laboratory, Dongguan, Guangdong 523808, China}
\author{Xiaobin Chen}
\email{chenxiaobin@hit.edu.cn}
\affiliation{School of Science and State Key Laboratory on Tunable Laser
  Technology and Ministry of Industry and Information Technology Key Lab of 
  Micro-Nano Optoelectronic Information System, Harbin Institute of Technology,
  Shenzhen 518055, China }
\author{Jia-Wei Mei}
 \email{meijw@sustech.edu.cn}
 \affiliation{Department of Physics, Southern University of Science and
   Technology, Shenzhen, 518055, China}
 \affiliation{Shenzhen Institute for Quantum Science and Engineering, Southern University of Science and Technology, Shenzhen 518055, China}
\affiliation{Shenzhen Key Laboratory of Advanced Quantum Functional Materials
   and Devices, Southern University of Science and Technology,    Shenzhen 518055, China}
\date{\today}

\begin{abstract}
In our study, we conduct magnetization and heat capacity measurements to investigate field-induced magnetic phase transitions within the newly synthesized compound \ce{K2Ni2(SeO3)3}, a spin-1 dimer system arranged on a triangular lattice. From our first-principles simulations, we determine that the spin system in \ce{K2Ni2(SeO3)3} can be represented as a two-dimensional triangular-lattice spin-1 dimer model, including an intra-dimer exchange of $J_1=0.32$~meV, an inter-dimer exchange of $J_2=0.79$~meV, and an easy-axis anisotropy of $D=0.14$~meV. The presence of easy-axis magnetic anisotropy explains the distinct magnetic phase diagrams observed under $c$-axis directional and in-plane magnetic fields. Notably, our investigation unveils a two-step phase transition with the magnetic field aligned with the $c$ direction. 
Our findings yield valuable insights into the magnetic phase transitions inherent to geometrically frustrated magnetic systems featuring dimer structures. 
\end{abstract}
\maketitle

\section{introduction}

The exploration of phase transitions holds profound significance in the realm of physics. These transitions reveal the fascinating world of symmetry breaking~\cite{Landau1937}, a powerful concept that aids in organizing our comprehension of the fundamental laws governing the universe~\cite{Coleman1985}. As time has progressed, our understanding of phase transitions has evolved significantly, introducing concepts like ``categorical symmetry'' through advancements in mathematics and physics~\cite{Kong2018,Kong2020,Kong2021,Ji2020,Chatterjee2023}.

In recent years, there has been a surge in interest in investigating field-induced magnetic phase transitions within quantum frustrated magnetic systems~\cite{Ono2003,Samulon2009,Samulon2008,Samulon2010,Shirata2012,Bordelon2019,Sheng2022,Sheng2023,Seabra2011,Yamamoto2014,Yamamoto2019,Gvozdikova2011,Kermarrec2021,Baek2017,Smith2023,Cui2023}. These systems, marked by geometric frustration, exhibit ground states  characterized by multiple degenerate configurations and strong quantum fluctuations~\cite{Balents2010,Moessner2006}. Consequently, they can manifest various distinct magnetic ground states when subjected to magnetic fields.  Besides the Bose–Einstein condensation (BEC) of diluted magnons near critical magnetic fields~\cite{Zapf2014,Sheng2022}, fractional magnetization plateaus are also observed in frustrated antiferromagnets. For instance, the 1/3-magnetization plateau appears in triangular compounds like Cs$_2$CuBr$_4$, Ba$_3$CoSb$_2$O$_9$, and Na$_2$BaCo(PO$_4$)$_2$~\cite{Ono2003,Ono2007,Alicea2009,Shirata2012,Sheng2022,Sheng2023}, and in the kagome antiferromagnet Li$_9$Fe$_3$(P$_2$O$_7$)$_3$(PO$_4$)$_2$ \cite{Kermarrec2021}. A recently discovered 1/9-magnetization plateau has been observed in the kagome antiferromagnet YCu$_3$(OD)$_{6+x}$Br$_{3-x}$ ($x\simeq0.5$)~\cite{Jeon2024}. Spin dimers can further enhance frustration in triangular antiferromagnets, leading to rich magnetic phase diagrams as seen in \ce{Ba3Mn2O8}~\cite{Samulon2009, Stone2008, Samulon2008, Samulon2010, TsuJii2005}. Therefore, the study of materials featuring frustrated magnetism in the presence of magnetic fields provides a unique and intricate context for delving into magnetic phase transitions. 

In this study, we explore magnetic phase transitions within the newly synthesized compound \ce{K2Ni2(SeO3)3}, which shares its isostructural nature with the sister compound \ce{K2Co2(SeO3)3}~\cite{Wildner1994, Zhong2020}. This compound presents a triangular lattice framework hosting a spin-1 Ni-Ni dimer system.   Combining first-principles simulations with our analysis, we conclude that the spin system in \ce{K2Ni2(SeO3)3} can be characterized as a two-dimensional spin-1 model. This model includes an intra-dimer exchange of $J_1=0.32$~meV, an inter-dimer exchange of $J_2=0.79$~meV, and an easy-axis anisotropy of $D=0.14$~meV. Our investigation probes the magnetic phase transitions in K$_2$Ni$_2$(SeO$_3$)$_3$ when subjected to magnetic fields applied both in-plane and out-of-plane. Notably, we uncover a successive two-step phase transition induced by fields for $B\|c$. 
The phase transition with the in-plane field $B\|ab$ is also studied, and exhibits distinct behavior compared to the case with $B\|c$, due to the presence of easy-axis magnetic anisotropy.

The subsequent sections of this paper are structured as follows. Section~\ref{sec:methods} provides details of our experimental methods, encompassing sample synthesis, sample characterization, and magnetization as well as heat capacity measurements. Additionally, we provide the theoretical setup for first-principles simulations. Moving forward to Section~\ref{sec:results}, we unveil the main outcomes of our investigation. In Section~\ref{sec:sample_detail}, we delve into the crystal structure and thermodynamic properties. Section~\ref{sec:ex_term} takes us into the realm of estimating exchange interactions within K$_2$Ni$_2$(SeO$_3$)$_3$ by an analysis that combines Curie-Weiss fitting of magnetic susceptibility data with the first-principles simulations. In Section~\ref{sec:fimt}, which is also the most significant part, we explore field-induced magnetic phase transitions for both the $c$-axis directional and in-plane magnetic fields. Finally, we summarize our results in Section~\ref{sec:sum}. 

\section{Methods}\label{sec:methods}
The synthesis of \ce{K2Ni2(SeO3)3} single crystals was carried out using the flux method, a procedure akin to that for \ce{K2Co2(SeO3)3}\cite{Zhong2020}. The initial materials, comprising NiO (Alfa Aesar, Ni: 78.5\%), KOH (Alfa Aesar, 99.98\%), and SeO$_2$ (Aladdin, 99.9\%), were mixed in a molar ratio of $1:4.8:4.8$ in preparation. In the glovebox, the mixed materials were  ground for 5 minutes and then transferred into an alumina crucible, which was subsequently placed inside a quartz tube. The tube was sealed under a vacuum pressure of 10$^{-3}$~Pa. This assembly was subjected to a heating process of 4~hours to 700~\celsius, followed by an 8~hour hold at that temperature. Subsequently, it was cooled over 100~hours to 200~\celsius~ maintained for an additional 4~hours, after which it was gradually cooled to room temperature. Finally, immersing the resulting materials in de-ionized water yielded transparent yellow single crystals, as depicted in the inset of Fig.~\ref{fig:figure2}(b). 

The crystal structure of \ce{K2Ni2(SeO3)3} was determined using single crystal X-ray diffraction (SCXRD) at 299~K,  with the Bruker D8 VENTURE diffractometer equipped with a PHOTON III C14 detector and the SCXRD using graphite-monochromatized $\lambda = 0.71073$~\AA~ Mo K$\alpha$ radiation. The collected data were used to decide the unit cell determination, data integration, and absorption correction. The raw data were corrected using APEX4 software. The structure was solved by the intrinsic phasing method with the SHELXT structure solution program1 and further refined using the ShelXL least squares refinement package~\cite{refinement} within the Olex-2 program~\cite{refinement2}. There was no higher symmetries for the compound verified by PLATON~\cite{Dolomanov2009}. Magnetization and heat capacity measurements were conducted utilizing the Quantum Design Magnetic Property Measurement System and Physical Property Measurement System, respectively. 

First-principles calculations were performed using the Vienna Ab-initio Simulation Package (VASP) with projected-augmented wave (PAW) potentials~\cite{Kresse1996, Blochi1994}. The revised Perdew–Burke–Ernzerhof parametrization (PBEsol) of the generalized gradient approximation was utilized for the exchange-correlation interaction~\cite{Perdew1996,Perdew2008}. Additionally, the LDA+U approach \cite{Dudarev1998} was employed with an effective Coulomb repulsion parameter $U_{\rm eff}=U-J_H=8$ eV. The energy tolerance was set to $1\times10^{-8}$ eV, and the energy cutoff was set at 520 eV. Lattice parameters were obtained from crystallographic data detailed in Tables~\ref{tab:table2} and Section~\ref{sec:results}. The initial positions of Se(\uppercase\expandafter{\romannumeral2}) atoms in a unit cell were determined as 0.78411 and 0.21589 in direct coordinates, with the randomness of Se atoms disregarded. A $9\times9\times2$ K-point mesh was utilized for optimizing the unit cell structure, with a force tolerance of $1\times10^{-5}$ eV/\AA~applied to each atom. The calculations were conducted with ferromagnetic spin configurations along the $c$-axis and included spin-orbit coupling~\cite{Steiner2016}. Initially, we assessed the magnetic anisotropy due to the spin-orbit effect in the unit cell, finding it to be small compared to the measured Curie-Weiss temperature obtained from magnetization measurements. Therefore, we did not include spin-orbit coupling in the self-consistent collinear spin calculation of the supercell structures. For the calculation of magnetic exchanges, a 2$\times$2$\times$1 supercell structure was used, along with a corresponding 5$\times$5$\times$2 K-point mesh.


\begin{figure*}[t]
	\centering
	\includegraphics[width=1.5\columnwidth]{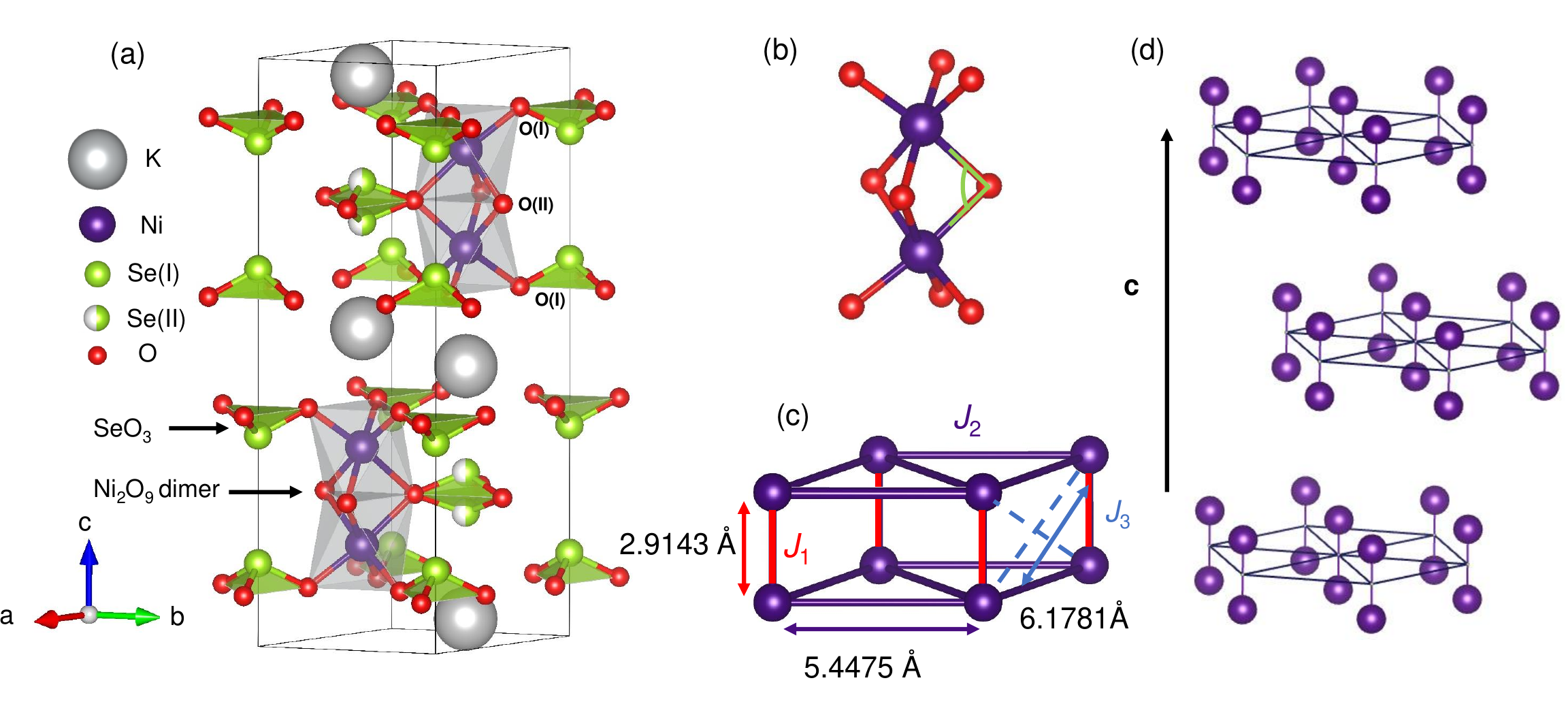}
	\caption{Crystal structure of \ce{K2Ni2(SeO3)3}. (a) The unit cell comprises two layers of \ce{Ni2O9} dimers (gray polyhedra).(b)The light green solid lines indicate Ni-O(\uppercase\expandafter{\romannumeral2})-Ni bond. (c) Schematic representation of nearest-neighbor ($J_1$), next-nearest-neighbor ($J_2$) and third-nearest-neighbor ($J_3$) exchange interactions. (d) Side view of the triangular lattice of Ni$^{2+}$-Ni$^{2+}$ dimers, stacking in an ABAB arrangement along the $c$-axis.} 
	\label{fig:figure1}
\end{figure*}
\section{results}\label{sec:results}

{\begin{table}[t]
	\caption{\centering{Crystal data and structure refinement for \ce{K2Ni2(SeO3)3}.}}
	\label{tab:table1}
 	 \centering
 \renewcommand\arraystretch{1.3}			
  \begin{tabular}{ll}
  \hline\hline
  Empirical formula & \ce{K2Ni2(SeO3)3}  \\
  \hline
  Temperature/K & 299 \\
  Formula weight & 576.46 \\
  Crystal system & Hexagonal \\
  Space group & $P6_{3}/mmc$ \\
  a/\AA & 5.4475(3) \\
  c/\AA & 17.4865(15) \\
  Volume/\AA$^3$ & 449.40(6) \\
  Z & 2 \\
  $\rho_{calc}$ (g/cm$^3$) & 4.260 \\
  $\mu$ (mm$^{-1}$) & 17.30 \\
  $F$(000) & 536 \\
  Crystal size/mm$^3$ & 0.05 $\times$ 0.04 $\times$ 0.02 \\
  Radiation & Mo $K{\alpha}$, $\lambda$ = 0.71073 \AA \\
  $T_{min}, T_{max}$ & 0.510, 0.746 \\
  $2\theta$ range for data collection/$^\circ$ & 4.475 to 29.715 \\
  Index ranges & $-7  \le h \le 7$ \\
   			& $-7  \le k  \le 7$  \\
			& $-24  \le l  \le 24$ \\
  Reflections collected & 8015 \\
  Independent reflections & 306 \\
  $R_{int}$ & 0.064 \\
  $R[F^2 > 2 \sigma (F^2)], wR(F^2), S$ & 0.030, 0.062, 1.18 \\
  \hline\hline
  \end{tabular}
\end{table}

\begin{table}[b]
	\caption{\centering{Crystallographic data in \ce{K2Ni2(SeO3)3}.}}
	\label{tab:table2}
  \centering
  \renewcommand\arraystretch{1.3}
  \begin{tabular}{cccccc}
  \hline\hline
    Atom & Wyckoff site & $x$ & $y$ & $z$ & Occupancy \\
    \hline
    K(\uppercase\expandafter{\romannumeral1}) & $4f$ & 0.33333 & 0.66667 & 0.0349(5) & 1 \\
    Ni(\uppercase\expandafter{\romannumeral1}) & $4f$ & 0.33333 & 0.66667 & 0.6666(7) & 1 \\
    Se(\uppercase\expandafter{\romannumeral1}) & $4e$ & 0 & 0 & 0.1413(4) & 1 \\
    Se(\uppercase\expandafter{\romannumeral2}) & $4f$ & 0.33333 & 0.66667 & 0.2158(9) & 0.5 \\
    O(\uppercase\expandafter{\romannumeral1}) & $12k$ & 0.16010 & 0.83990 & 0.5973(8) & 1 \\
    O(\uppercase\expandafter{\romannumeral2}) & $6h$ & 0.49930 & 0.50070 & 0.2500 & 1 \\
    \hline\hline
  \end{tabular}
\end{table}

\subsection{Sample characterization}\label{sec:sample_detail}
Tables~\ref{tab:table1} and \ref{tab:table2} present detailed crystal structure information for K$_2$Ni$_2$(SeO$_3$)$_3$. K$_2$Ni$_2$(SeO$_3$)$_3$ is isostructural with its sister compound \ce{K2Co2(SeO3)3}~\cite{Wildner1994,Zhong2020}, and crystallizes in the hexagonal space group $P6_{3}/mmc$ (No.~194) with the lattice parameters of $a = b = 5.4475(3)$~\AA, $c = 17.4865(15)$~\AA. While K(\uppercase\expandafter{\romannumeral1}), Ni(\uppercase\expandafter{\romannumeral1}), O(\uppercase\expandafter{\romannumeral1}), O(\uppercase\expandafter{\romannumeral2}) and Se(\uppercase\expandafter{\romannumeral1}) atoms fully occupy crystallographic positions, the Se(\uppercase\expandafter{\romannumeral2}) atoms, located on the Wyckoff position $4f$, exhibit a split into two sites with equal occupancy. The structural disorder due to the Se(\uppercase\expandafter{\romannumeral2}) random occupancy is the same as that in K$_2$Co$_2$(SeO$_3$)$_3$, which is thoroughly discussed in Ref.~\cite{Zhong2020}.

Figure~\ref{fig:figure1} schematically shows the crystal structure of \ce{K2Ni2(SeO3)3}, where two NiO$_6$ octahedra share a common O$_3$-triangle face, forming a Ni$_2$O$_9$ dimer. The NiO$_6$ octahedron exhibits a trigonal elongation along the $c$-axis, resulting in easy-axis magnetic anisotropy. Within the Ni$_2$O$_9$ dimer, the bond angle $\angle$Ni-O(\uppercase\expandafter{\romannumeral2})-Ni(Green lines in Fig.~\ref{fig:figure1}(b)) between Ni$^{2+}$ ions at the bridging O(\uppercase\expandafter{\romannumeral2}) atoms measures 85.4\degree, close to 90\degree.
Consequently, Ni$_2$O$_9$ constitutes an easy-axis spin-1 dimer. The dimers are interconnected through Se(\uppercase\expandafter{\romannumeral1},\uppercase\expandafter{\romannumeral2})O$_3$ tripods. While the Se(\uppercase\expandafter{\romannumeral1})O$_3$ tripods establish connections between the Ni$_2$O$_9$ dimers through the oxygen atoms located on the upper and lower O$_3$-triangles, the Se(\uppercase\expandafter{\romannumeral2})O$_3$ tripods form connections by utilizing the oxygen atoms shared on the common middle O$_3$-triangle face.

\begin{figure}[t]
  \centering  
  \includegraphics[width=\columnwidth]{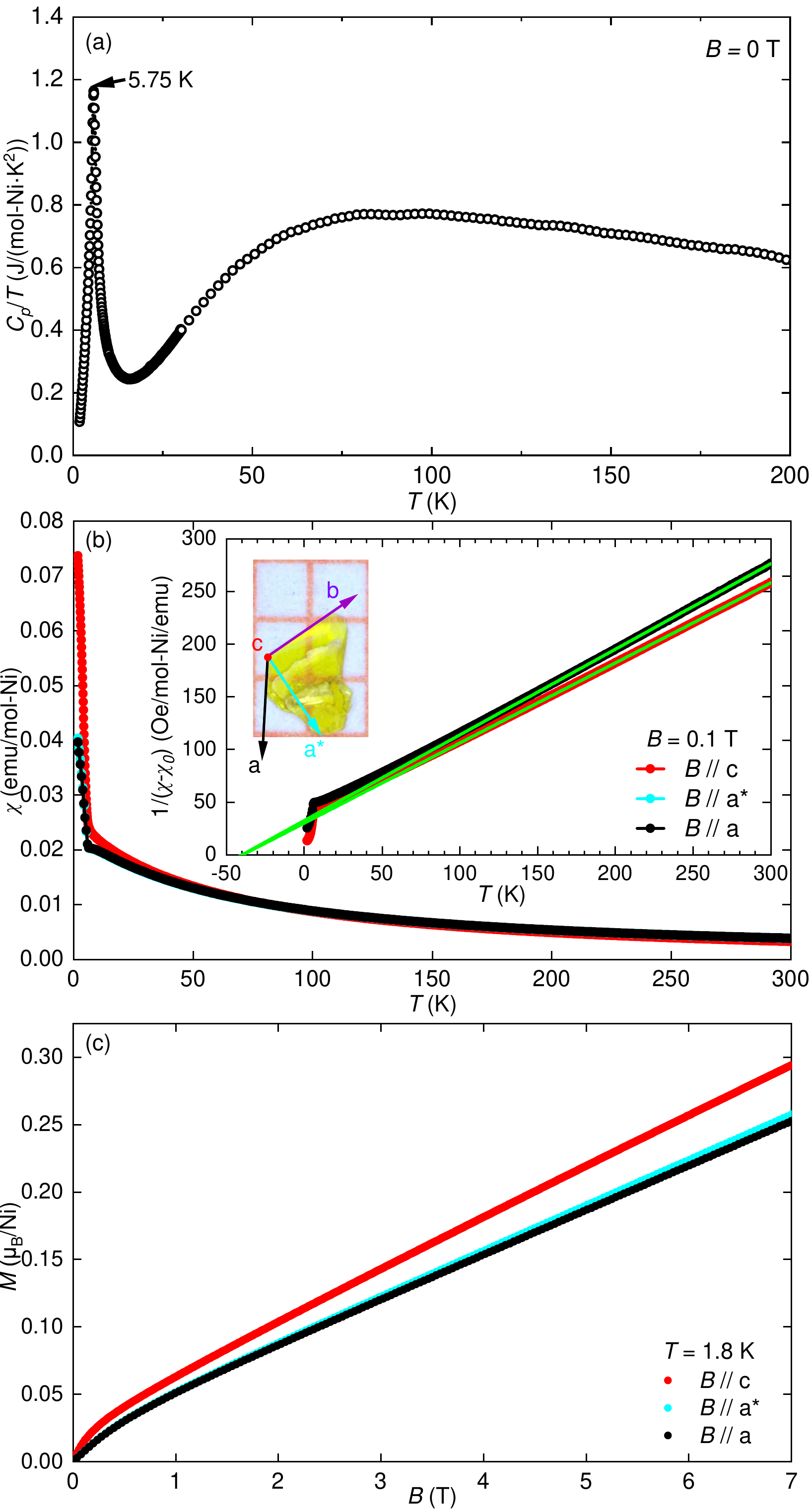}
  \caption{(a) Zero-field specific heat. (b) Temperature dependent magnetic susceptibility with $B=0.1$~T. Inset is the inverse magnetic susceptibility 1/($\chi-\chi_{0}$)(\emph{T}) and the green lines are the Curie-Weiss fittings. (c) Field dependent magnetization at 1.8~K.}
  \label{fig:figure2}
\end{figure}
Figure~\ref{fig:figure2} presents the basic thermodynamic properties of K$_2$Ni$_2$(SeO$_3$)$_3$. The temperature-dependent zero-field heat capacity divided by temperature $C_{p}(T)/T$ in Fig.~\ref{fig:figure2}(a), and the {magnetic susceptibility $\chi(T)$} with an applied magnetic field of $B=0.1$~T in Fig.~\ref{fig:figure2}(b), reveal a well-defined manetic phase transition occurring at the critical temperature $T_c=5.75$~K. No additional discernible thermodynamic anomalies are observed above $T_c$ in either $C_{p}(T)$ or $\chi(T)$. Below $T_c$, themagnetic susceptibility $\chi$ exhibits a larger magnitude when subjected to a $c$-axis directional field in comparison to an in-plane field, thereby indicating the easy-axis magnetic anisotropy, which is further confirmed by the field dependent magnetization $M(B)$ at 1.8~K as shown in Fig.~\ref{fig:figure2}(c).

\subsection{Exchange interactions}\label{sec:ex_term} In \ce{K2Ni2(SeO3)3}, the Ni$^{2+}$ ions possess a spin-1 $3d^{8}$ electronic configuration. Within a Ni$_2$O$_9$ dimer, two Ni$^{2+}$ ions interact magnetically through the three shared oxygen atoms, resulting in the intra-dimer exchange interaction ($J_1$ in Fig. \ref{fig:figure1}(c)). According to the Goodenough-Kanamori rule~\cite{Goodenough2008}, the Ni-O(\uppercase\expandafter{\romannumeral2})-Ni super-exchange pathways with bond angle $\angle$Ni-O(\uppercase\expandafter{\romannumeral2})-Ni of 85.4\degree~ lead to a ferromagnetic interaction with $J_{1,{\rm super-exchange}}<0$.  However, the direct exchange interaction leads to an antiferromagnetic interaction with $J_{1,{\rm exchange}}>0$. Therefore, the total intra-dimer exchange interaction $J_1=J_{1,{\rm super-exchange}}+J_{1,{\rm exchange}}$ requires further estimation.

Within the triangular lattice of Ni$_2$O$_9$ dimers, the inter-dimer magnetic interactions, denoted as $J_2$ and $J_3$ in Fig.~\ref{fig:figure1}(c), are mediated through two distinct Se(\uppercase\expandafter{\romannumeral1},\uppercase\expandafter{\romannumeral2})O$_3$ tripods, involving the Ni-O(\uppercase\expandafter{\romannumeral1})-Se(\uppercase\expandafter{\romannumeral1})-O(\uppercase\expandafter{\romannumeral1})-Ni and Ni-O(\uppercase\expandafter{\romannumeral2})-Se(\uppercase\expandafter{\romannumeral2})-O(\uppercase\expandafter{\romannumeral2})-Ni paths, respectively. Se(\uppercase\expandafter{\romannumeral1})O$_3$ facilitates the super-exchange interaction $J_2$, while Se(\uppercase\expandafter{\romannumeral2})O$_3$ plays a role in both $J_2$ and $J_3$. It's worth noting that the random occupation of Se(\uppercase\expandafter{\romannumeral2}) could potentially induce disorder effects to $J_2$ and $J_3$. However, the exploration of these disorder effects is beyond the scope of the present study, and thus we do not consider them in this paper.


The triangular-lattice layer of Ni$_2$O$_9$ dimers is effectively isolated from each other by non-magnetic K$^{+}$ ions, resulting in minimal inter-layer interactions. Consequently, the spin system within K$_2$Ni$_2$(SeO$_3$)$_3$ can be described as a two-dimensional spin-1 dimer model expressed by the following Hamiltonian 
\begin{eqnarray}
 	\label{eq:3}
 	H &=& J_1\sum_i {\bf S}_{i1}\cdot{\bf S}_{i2}+\sum_{\langle ij
  \rangle}\big( J_2({\bf S}_{i1}\cdot{\bf S}_{j1}+{\bf S}_{i2}\cdot{\bf
  S}_{j2})\\
    &+&\big.J_3({\bf S}_{i1}\cdot{\bf S}_{j2}+{\bf S}_{i2}\cdot{\bf S}_{j1})\big)-D\sum_{i}\big((S_{i1}^z)^2+(S_{i2}^z)^2 \big).\nonumber
\end{eqnarray}
Here, ${\bf S}_{i,1/2}$ represents the spin-1 operator of the first/second Ni$^{2+}$ ion within the Ni$_2$O$_9$ dimer on the $i$-th site, and $\langle ij \rangle$ denotes the nearest-neighbor bond for the dimers. The parameters $J_{1}$, $J_{2}$, and $J_{3}$ correspond to the exchange interactions as illustrated in Fig.~\ref{fig:figure1}(b), while $D$ accounts for the easy-axis magnetic anisotropy.

\begin{figure}[b]
  \centering
  \includegraphics[width=\columnwidth]{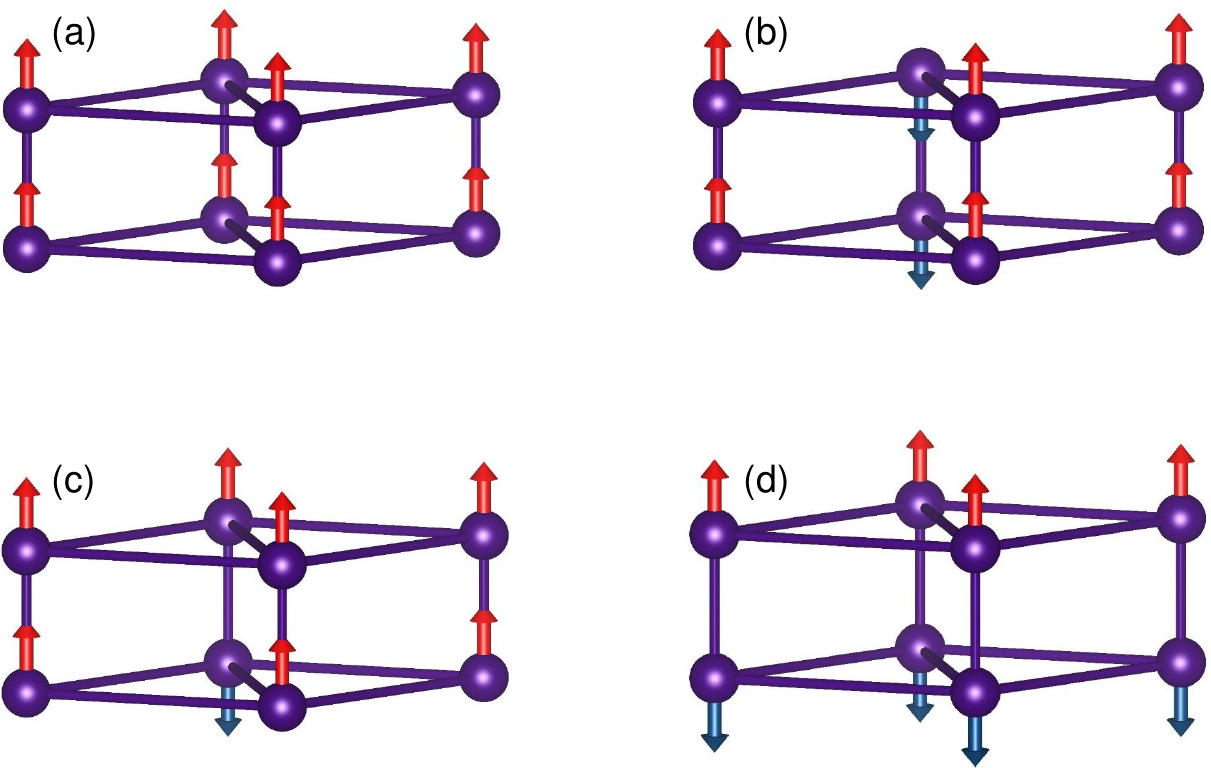}
  \caption{Four different spin configurations in the first-principles simulation. 
    The arrows indicate spin directions on the Ni$^{2+}$ irons.}
  \label{fig:figure3}
\end{figure}

To provide a preliminary estimate of the magnetic anisotropy parameter $D$, we compare the magnetization magnitudes of K$_2$Ni$_2$(SeO$_3$)$_3$ for both $c$-axis directional and in-plane magnetic fields ($B//a$, $B//a^{*}$) at 1.8~K in Fig.~\ref{fig:figure2}(c). For field in the ab plane, the magnetic susceptibility is nearly isotropic. We find that $(M_c-M_{ab})/M_{ab}$$ \simeq0.2$ under $B=7$~T. This provides a rough determination of the magnetic anisotropy $D$, which is on the order of 10\% of the interactions between the Ni$_2$O$_9$ dimers. The magnetic anisotropy plays a role in the different field-induced phase diagrams for the $c$-axis directional and in-plane magnetic fields, which we will delve into further in Figures~\ref{fig:figure8} and \ref{fig:figure11}. 

To determine magnetic interactions in K$_2$Ni$_2$(SeO$_3$)$_3$, we performed fittings on the temperature-dependent magnetic susceptibility shown in Fig.~\ref{fig:figure2}(b). These fittings are conducted by utilizing the Curie-Weiss law ${\chi}=\frac{C}{T-\Theta}+\chi_0$ within the temperature range of 200~K to 300~K, where $C$ is the Curie constant, $\Theta$ is the Curie-Weiss temperature and $\chi_0$ denotes the $T$-independent contribution. The fittings yield values of $\Theta_{c}={{-39.77 \pm 0.95}}$~K and $\Theta_{ab}={{-40.12\pm 1.43}}$~K, indicating the overall antiferromagnetic interactions in K$_2$Ni$_2$(SeO$_3$)$_3$. Additionally, we determine the $g$-factors, with $g_{c}$ = ${{2.30 \pm 0.01}}$, $g_{ab}$ = ${{2.22 \pm 0.01}}$ 

The Curie-Weiss fitting, though informative, does not provide separate values for $J_1$, $J_2$, and $J_3$. To obtain these individual exchange parameters, we turn to first-principles simulations, which offer a more detailed insight into the magnetic interactions. We initially compared the total energies for spins along the $a$ and $c$-axes to estimate the magnetic anisotropy as $D=0.14$ meV. Additionally, we estimated the inter-dimer interaction along the $c$-axis to be $J'=0.03$ meV, confirming the dimensionality of 2 for the spin system in \ce{K2Ni2(SeO3)3}.

In our simulations, we considered four distinct magnetization configurations, as depicted in Figure~\ref{fig:figure3}, and calculated their corresponding total energies. The two dimer layers in the unit cell were set to have the same spin configurations. Considering the Hamiltonian in Eq.~(\ref{eq:3}), the total energies for the four spin configurations are as follows:
\begin{eqnarray}
  E_a&=&E_0+8J_1+48J_2+48J_3\nonumber\\
  E_b&=&E_0+8J_1\nonumber\\
  E_c&=&E_0+4J_1+24J_2+24J_3\nonumber\\
  E_d&=&E_0-8J_1+48J_2-48J_3.
\end{eqnarray}
Our theoretical calculations yielded $J_{1}=0.32$ meV, $J_{2}=0.79$ meV, and $J_{3}=0.01$ meV. These values, along with $D=0.14$ meV obtained through simulations, complement our experimental findings and provide a more comprehensive understanding of the magnetic interactions. The theoretical Curie-Weiss temperature calculated using these parameters, approximately $\Theta=-2(J_1+6J_2+6J_3)/3\simeq-40$ K, closely aligns with the value derived from the experimental fitting, reinforcing the consistency of our results.}

\subsection{Field-induced magnetic transitions}\label{sec:fimt}

\subsubsection{$B\|c$}\label{sec:THc}
\begin{figure}[t]
  \centering  
  \includegraphics[width=0.9\columnwidth]{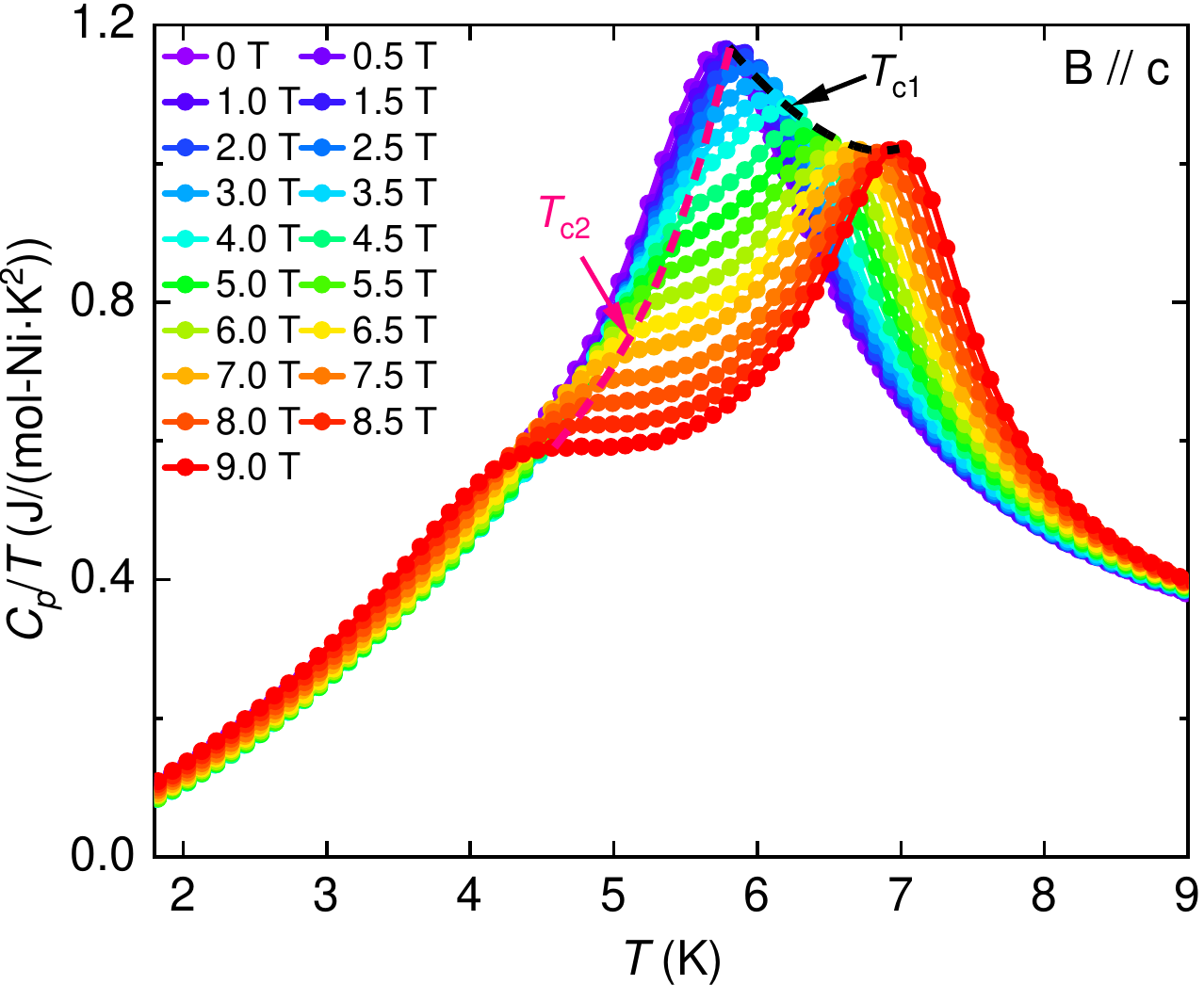}
  \caption{Specific heat $C_p(T)/T$ and  successive phase
    transition temperatures $T_{c1}$ and $T_{c2}$ indicated by dashed lines under different fields $B\|c$. }
  \label{fig:figure4}
\end{figure}
\begin{figure*}[t] \centering
  \includegraphics[scale=0.6]{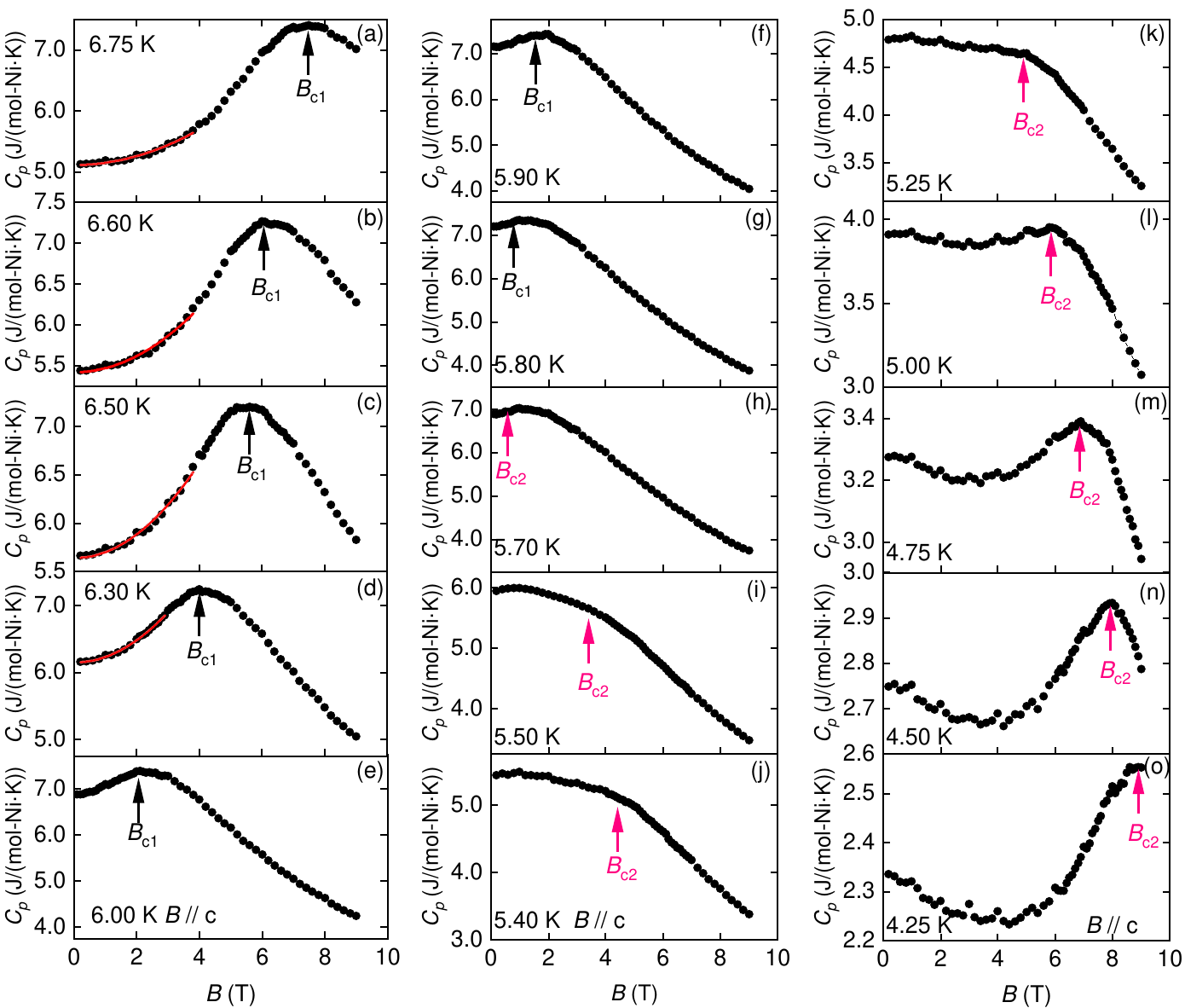}
  \caption{Field-dependent specific heat $C_p(B)$ and critical magnetic fields $B_{c1}$ and $B_{c2}$ at selected temperatures with $B\|c$. The solid red lines in (a)-(d) represent the fitting results of $\Delta C_p(B) \propto B^2$.}
  \label{fig:figure5}
\end{figure*}
We can only resolve a single magnetic phase transition occurring at $T_c=5.75$~K in the zero-field heat capacity measurement displayed in Fig.~\ref{fig:figure2}(a). The introduction of a magnetic field prompts us to explore the field-induced magnetic phase transition, which is depicted in the temperature-dependent specific heat divided by temperature $C_p(T)/T$ for \ce{K2Ni2(SeO3)3} with the magnetic field aligned along the $c$-axis, as presented in Figure~\ref{fig:figure4}. As the applied magnetic field strength increases, the original transition at $T_c$ gradually splits into two distinct transitions, labeled as $T_{c1}$ and $T_{c2}$. Notably, these transition points, $T_{c1}$ and $T_{c2}$, exhibit distinct behaviors with respect to the magnetic field strength. The black dashed line in Fig.~\ref{fig:figure4} signifies the shift of $T_{c1}$ to higher temperatures as the magnetic field is increased, whereas the pink dashed line indicates the opposite trend for $T_{c2}$, shifting to lower temperatures. It is important to observe that while the peak at $T_{c1}$ retains its sharpness, resembling the zero-field transition, the transition peak at $T_{c2}$ exhibits a significantly broader profile. This contrasting behavior in the peak profiles highlights the distinct nature of these phase transitions occurring at $T_{c1}$ and $T_{c2}$. 

The heat capacity measurements in Fig.~\ref{fig:figure4} provide compelling evidence for the existence of two successive transitions, and hence the presence of an intermediate phase. This intermediate state is anticipated to manifest itself in the field-dependent specific heat $C_p(B)$ when probed at various fixed temperatures.  As the applied magnetic field induces a progressive upward shift of $T_{c1}(B)$, as indicated by the black dashed line in Fig.~\ref{fig:figure4}, one would expect to identify the same phase transition point $B_{c1}(T)$ within $C_p(B)$ when maintaining a constant temperature above $T_c=5.75$~K. Likewise, keeping the temperature fixed below $T_c$ allows us to discern the phase transition occurring at $B_{c2}(T)$, which corresponds to $T_{c2}(B)$. To validate the existence of the intermediate phase, a comprehensive analysis of the detailed data for the field-dependent specific heat $C_p(B)$ is presented in Figure~\ref{fig:figure5}.

\begin{figure}[t] \centering
  \includegraphics[width=\columnwidth]{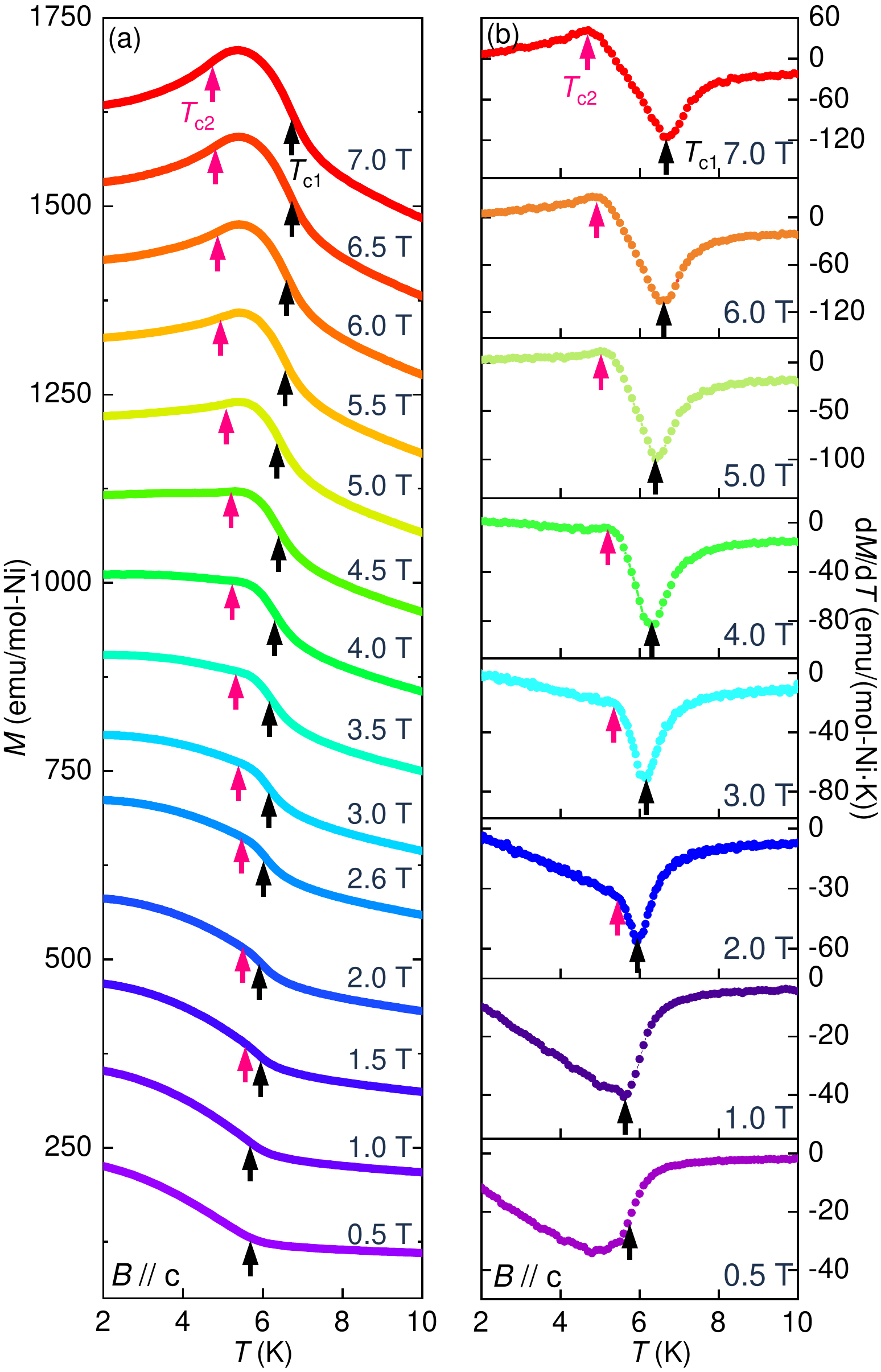}
  \caption{(a) Temperature-dependent magnetization with the magnetic field aligned along the $c$-axis. (b) Corresponding temperature derivative of the magnetization, $dM/dT$.}
  \label{fig:figure6}
\end{figure}

In Fig.~\ref{fig:figure5}(a), we examine the field-dependent specific heat $C_p(B)$ while maintaining a constant temperature of $T = 6.75$ K above $T_c$. At this temperature, K$_2$Ni$_2$(SeO$_3$)$_3$ is in a paramagnetic state under low magnetic fields. As the magnetic field $B$ increases, we observe an enhancement in $C_p(B)$, particularly at low fields. By applying polynomial fitting to the data within the range of 0~T to 4~T, we find that the increase in $C_p(B)$ approximately follows a relationship of $\Delta C_p(B) \propto B^2$, where $\Delta C_p(B)=C_p(B)-C_p(0)$. This behavior aligns with the field dependence of specific heat typically observed in paramagnetic states~\cite{VanVleck1937}. Upon reaching the critical value of $B_{c1} = 7.5$~T, a peak emerges in $C_p(B)$, indicating a transition into the intermediate state for the system. Similar trends are observed in Figs.~\ref{fig:figure5}(b)-(g) for temperatures above $T_c$. In this range, the critical magnetic field $B_{c1}(T)$ decreases as the temperatures at which the measurements are taken decrease.

Around $T_c = 5.75$ K (Figs.~\ref{fig:figure5}(f)-(k)), the transition peaks in $C_p(B)$ become notably broad and challenging to discern clearly. However, as we continue to lower the temperature below $T_c$, the transition peak in $C_p(B)$ becomes pronounced again, allowing us to confidently identify the critical field $B_{c2}(T)$. When we delve below 5.0~K in Figs.~\ref{fig:figure5}(l)-(o), a distinct pattern emerges in the field dependence in $C_p(B)$, differing from that in Figs.~\ref{fig:figure5}(a)-(f). Here, the specific heat $C_p(B)$ initially decreases as the field increases at low fields, then experiences an enhancement with further increases in $B$, and reaches a peak at $B_{c2}$ before transitioning into the intermediate phase. Notably, with decreasing holding temperatures, the critical field $B_{c2}(T)$ exhibits a consistent increase. The distinct trends observed in $B_{c1}(T)$ and $B_{c2}(T)$ underscore the intricate interplay between temperature and magnetic field, providing validation for the existence of the intermediate phase.

The successive phase transitions in K$_2$Ni$_2$(SeO$_3$)$_3$ are not only evident in the specific heat measurements (as seen in $C_p(T)$ in Fig.~\ref{fig:figure4} and $C_p(B)$ in Fig.~\ref{fig:figure5}) but also discernible in the magnetization data. Figure~\ref{fig:figure6} presents the temperature-dependent magnetization $M(T)$ and its associated differential magnetization concerning temperature ($dM/dT$) under varying magnetic fields. At lower magnetic field strengths, as the temperature decreases to 2 K, the magnetization in Fig.~\ref{fig:figure6}(a) exhibits a steady increase. Conversely, at higher magnetic fields, the magnetization initially ascends as the temperature decreases, but it subsequently begins to decline.

Notably, the temperature-dependent differential magnetization ($dM/dT$) in Fig.~\ref{fig:figure6}(b) reveals two distinctive kinks. These kink temperatures, indicated by arrows, agree with the critical temperatures $T_{c1}$ and $T_{c2}$ as determined through specific-heat measurements. A sudden increase in magnetization becomes evident at $T_{c1}$. While the rate of enhancement diminishes until reaching $T_{c2}$ at low magnetic fields, this upward trend in magnetization not only decelerates but also ultimately reverses, leading to a decrease in magnetization at high magnetic fields. 

\begin{figure}[t]
  \centering
  \includegraphics[width=0.9\columnwidth]{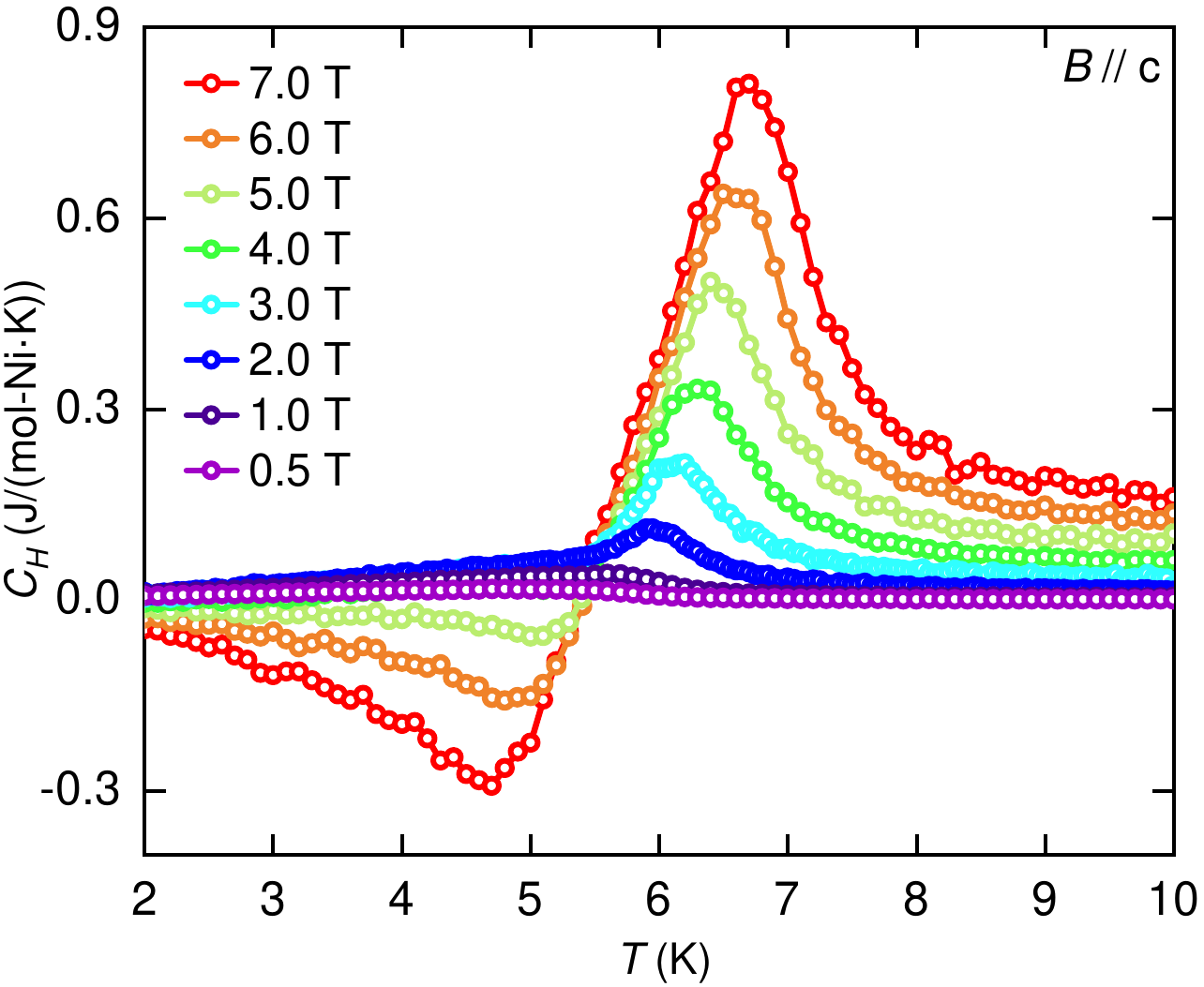}
  \caption{Heat capacity contribution $C_H=-\mu_0H(\frac{\partial M}{\partial T})_H$ from the work performed by the magnetic field on the magnetization.}
  \label{fig:figure7}
\end{figure}

To further underscore the coherent alignment between the heat capacity and magnetization measurements, we reconfigure the data by plotting $C_H=-\mu_0H(\frac{dM}{dT})_H$ in Figure~\ref{fig:figure7}. This quantity signifies the heat capacity contribution from the work performed by the magnetic field on the magnetization. According to the second law of thermodynamics,
\begin{eqnarray}
  \label{eq:2nd}
  dQ=dU-\mu_0HdM,
\end{eqnarray}
we know the specific heat at constant field
\begin{eqnarray}
  \label{eq:Cp}
  C=C_M+C_H,
\end{eqnarray}
where $C_M=(\frac{\partial U}{\partial T})_M$ represents the specific heat at a constant magnetization, derived from the internal energy $U$. Meanwhile $C_H=-\mu_0H(\frac{\partial M}{\partial T})_H$ accounts for the contribution arising from the work done by the magnetic field. It's important to note that when the magnetic field remains constant, the magnetization $M$ varies with temperature, and consequently, $C_M$ also depends on the magnetic field. Nonetheless, $C_H$ specifically represents how changes in magnetization influence the heat capacity due to the work done by the magnetic field.

\begin{figure}[b]
\centering  
\includegraphics[width=\columnwidth]{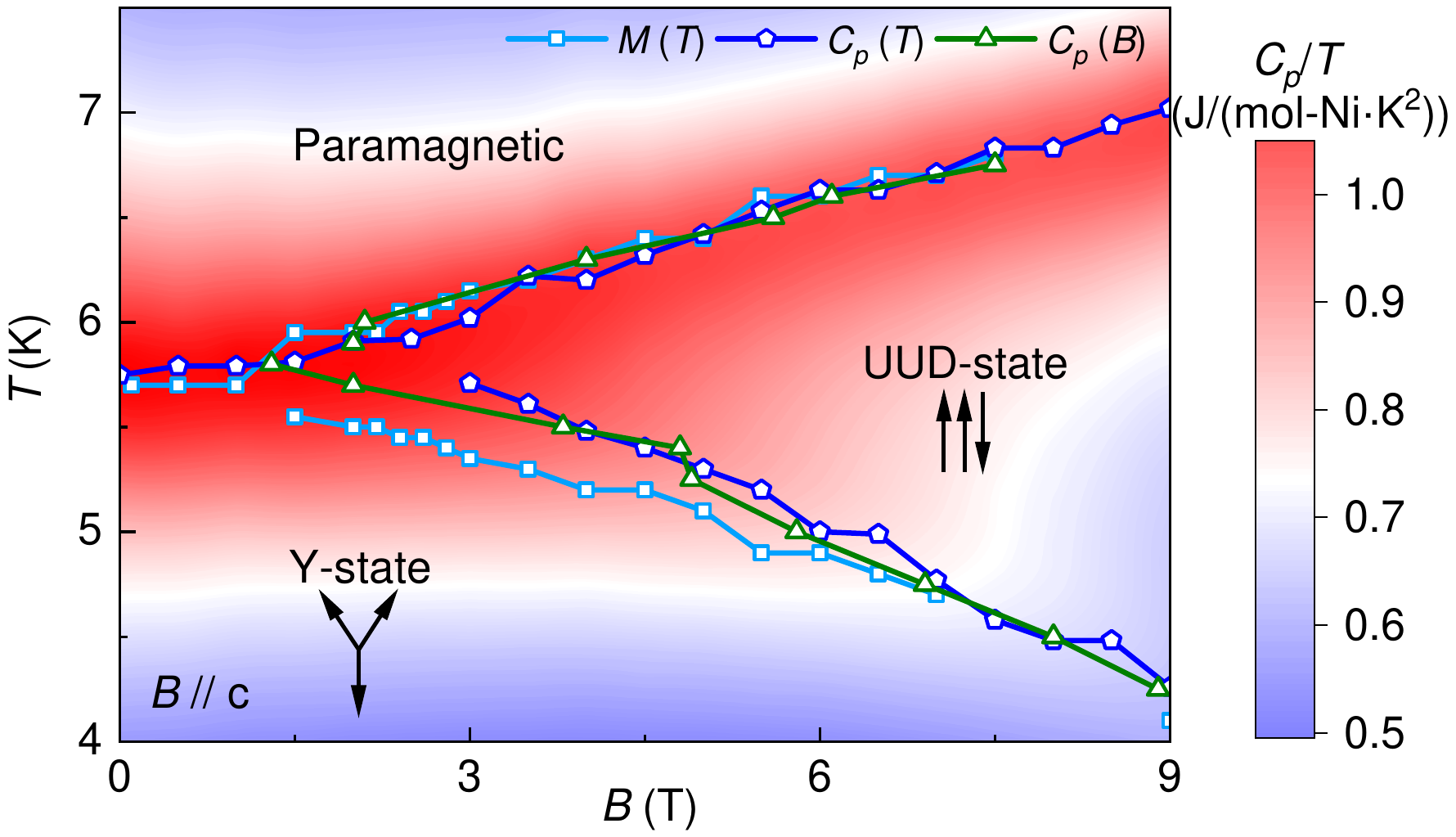}  
\caption{Magnetic phase diagram for $B \| c$ with the phase boundaries determined by the collected data from $C_p(T)$ in Fig.~\ref{fig:figure4}, $C_p(B)$ in Fig.~\ref{fig:figure5} and $M(T)$ in Fig.~\ref{fig:figure6}. }
\label{fig:figure8}
\end{figure}

\begin{figure}[t]
  \centering
\includegraphics[width=1\columnwidth]{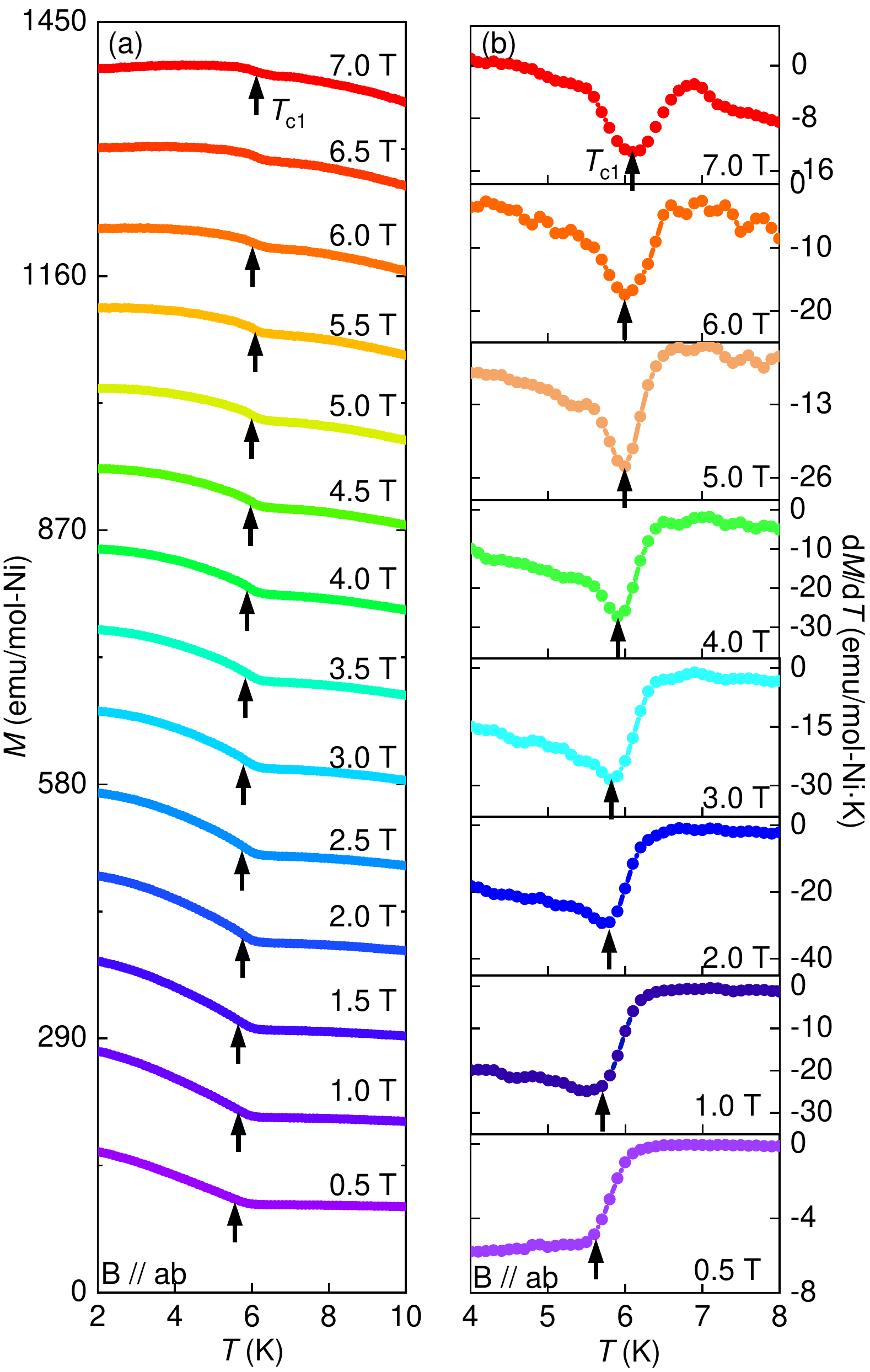}
\caption{(a) Temperature-dependent magnetization with the in-plane magnetic field. (b) Corresponding temperature derivative of the magnetization, $dM/dT$.}
    \label{fig:figure9}
\end{figure}
\begin{figure}[t]
\centering  
\includegraphics[width=0.9\columnwidth]{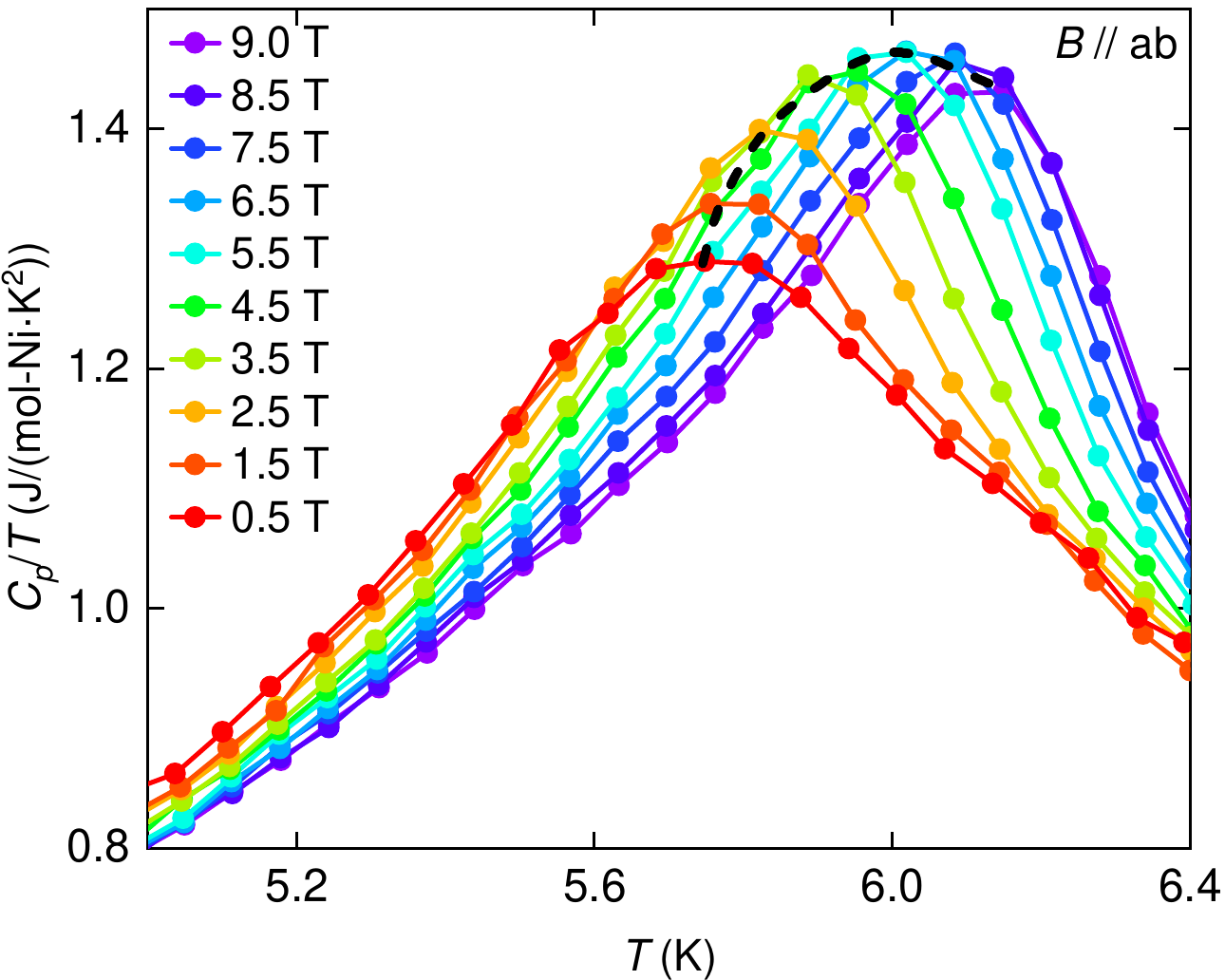}    
  \caption{Specific heat under different in-plane fields $B\|ab$.}
\label{fig:figure10}
\end{figure}
\begin{figure}[b]
  \centering  
  \includegraphics[width=\columnwidth]{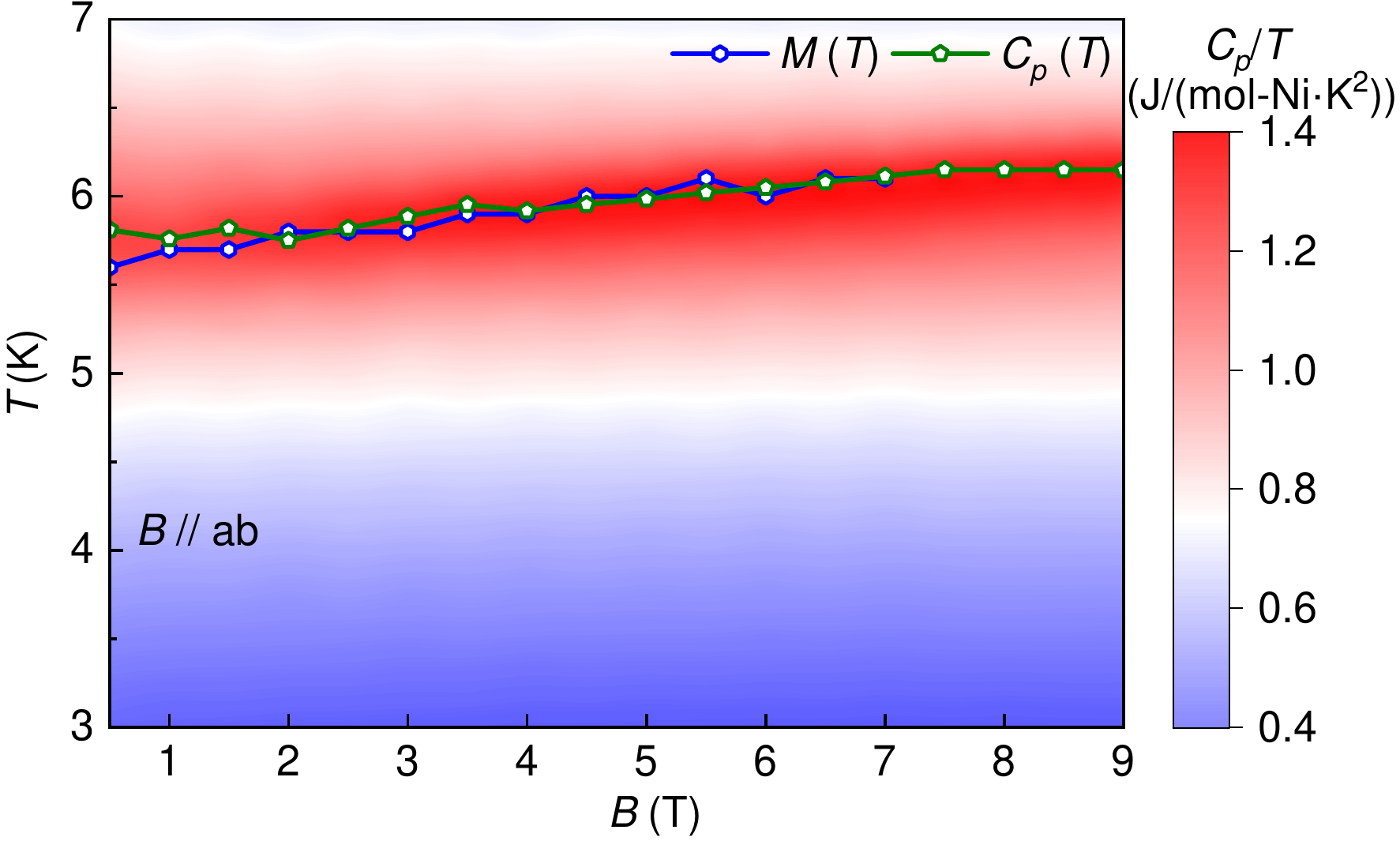}   
  \caption{Magnetic phase diagram for $B \| ab$.}
  \label{fig:figure11}
\end{figure}

Upon heat capacity $C_H$ arising from the magnetic field's work on the magnetization in Fig.~\ref{fig:figure7}, we conduct a comparative analysis by examining the temperature-specific heat $C_p(T)/T$ shown in Fig.~\ref{fig:figure4}. We observe that the first transition peak in $C_p(T)/T$ at $T_{c1}$ closely resembles the corresponding peak in $C_H$. This resemblance suggests that at the transition point $T_{c1}$, there is a substantial alteration in magnetization aligned with the field direction. However, the situation deviates significantly at the second transition $T_{c2}$. Here, the behavior of $C_H$ differs notably from that of $C_p(T)/T$ and $C_H$ even exhibits a negative value under strong magnetic fields. These observations imply that, although the magnetization along the field direction does not undergo significant changes at $T_{c2}$, the internal energy experiences substantial variations. This may indicate that the magnetization perpendicular to the field direction undergoes pronounced shifts during this phase transition. We can also draw parallels between the field-dependent heat capacity behavior $C_p(B)$ at low magnetic fields in Fig.~\ref{fig:figure5}, and the patterns observed in $C_H$ in Fig. fig:figure7. It becomes evident that as we keep the temperature constant and increase the magnetic field strength, both $C_H$ and $C_p(B)$ exhibit similar trends in their values at low fields.

Upon gathering critical points from the successive magnetic transitions $T_{c1}$ and $T_{c2}$ in $C_p(T)/T$ (Fig.~\ref{fig:figure4}), as well as $M(T)$ (Fig.~\ref{fig:figure6}), and $B_{c1}$ and $B_{c2}$ in $C_p(B)$ (Fig.~\ref{fig:figure5}), we can construct the magnetic phase diagram for $B\|c$, as illustrated in Figure~\ref{fig:figure8}. The color intensity in the diagram reflects the values of $C_p/T$, and it approximately aligns with the critical points. The phase diagram is effectively divided into three distinct phases.From our theoretical investigations, the magnetic exchanges in \ce{K2Ni2(SeO3)3} is dominated by $J_2$ and the spin system can be approximated by the spin-1 triangular antiferromagnet. Leveraging insights gained from both Monte Carlo simulations~\cite{Seabra2011,Gvozdikova2011,Melchy2009} and experimental studies\cite{Ishii2009,Lee2014} conducted on the traditional triangular-lattice antiferromagnet under magnetic fields, we propose a sequence of phase transitions. Initially, the system undergoes a continuous phase transition, transitioning from the paramagnetic state to the UUD state, which breaks the $Z_3$ lattice symmetry. As the temperature continues to decrease, a Berezinskii-Kosterlitz-Thouless phase transition ensues, leading the system into the ``Y state'' from the UUD state, which breaks the $c$-axis spin rotation symmetry. 
This effect of the $J_1$ in the phase transition requires more theoretical investigations for further validation.

\subsubsection{$B\|ab$}
To comprehensively explore the field-induced magnetic phases in K$_2$Ni$_2$(SeO$_3$)$_3$, we extend our investigations to encompass in-plane magnetic fields aligned with the crystallographic $ab$ plane ($B\|ab$). Following a similar interpretation approach as outlined in Section~\ref{sec:THc}, the magnetization data presented in Figure~\ref{fig:figure9} and the heat capacity measurements depicted in Figure~\ref{fig:figure10} enabled us to extract critical transition temperatures $T{c}(B)$ and fields $B{c}(T)$, respectively, under in-plane magnetic fields. These critical values are subsequently used to construct a magnetic phase diagram, which is presented in Figure~\ref{fig:figure11}. This phase diagram serves as a visual representation of how the magnetic phases evolve under the influence of in-plane magnetic fields. 

It is noteworthy that our findings reveal a notable disparity between the phase diagrams for $B\|c$ and $B\|ab$. This divergence can be attributed to the impact of on-site magnetic anisotropy, characterized by parameter $D$, which significantly influences the magnetic behavior of the Ni$^{2+}$ ions within the crystal structure of K$_2$Ni$_2$(SeO$_3$)$_3$. This observation underscores the critical importance of considering crystallographic orientations and the effects of magnetic anisotropy when investigating the magnetic properties of intricate materials such as K$_2$Ni$_2$(SeO$_3$)$_3$.

\section{Summary and conclusion}\label{sec:sum}
In our comprehensive investigation, we delved into the intriguing field-induced magnetic phase transitions within the newly synthesized compound \ce{K2Ni2(SeO3)3}, characterized by a spin-1 dimer system arranged on a triangular lattice. Through an integrated approach encompassing magnetization and heat capacity measurements, coupled with Curie-Weiss fitting of the magnetization data and first-principles simulations, we elucidated the underlying exchange interactions governing the behavior of the spin-1 Ni$^{2+}$ ions in K$_2$Ni$_2$(SeO$_3$)$_3$. 


One of the most notable findings of our investigation was the identification of a two-step phase transition when the magnetic field was aligned with the $c$ direction. The first transition, from a paramagnetic state to an UUD state, was characterized by the breaking of the $Z_3$ lattice symmetry. Subsequently, a Berezinskii-Kosterlitz-Thouless transition ensued, marked by the breaking of the c-axis spin-rotation symmetry, leading to the formation of what we term the ``Y state'' at low temperatures.

In conclusion, our study provides valuable insights into the intricate magnetic phase transitions inherent to geometrically frustrated magnetic systems featuring dimer structures. The newly synthesized compound \ce{K2Ni2(SeO3)3} serves as an intriguing model system, shedding light on the rich and complex behavior of spin-1 systems arranged on triangular lattices under the influence of magnetic fields. These findings not only expand our fundamental understanding of quantum magnetism but also hold promise for potential applications in emerging technologies.

\begin{acknowledgments}
  This work is supported by the National Key Research and Development Program of China (Grant No. 2021YFA1400400), the National Natural Science Foundation of China (Grant No.12204223), Shenzhen Fundamental Research Program (Grant No. JCYJ20220818100405013), the Guangdong Innovative and Entrepreneurial Research Team Program (Grants No. 2017ZT07C062), Shenzhen Key Laboratory of Advanced Quantum Functional Materials and Devices (Grant No. ZDSYS20190902092905285), Guangdong Basic and Applied Basic Research Foundation (Grant No. 2020B1515120100), Shenzhen Science and Technology Program (Grant No. RCYX20221008092848063).
\end{acknowledgments}
\bibliography{Ni2}

\begin{thebibliography}{47}%
\makeatletter
\providecommand \@ifxundefined [1]{%
 \@ifx{#1\undefined}
}%
\providecommand \@ifnum [1]{%
 \ifnum #1\expandafter \@firstoftwo
 \else \expandafter \@secondoftwo
 \fi
}%
\providecommand \@ifx [1]{%
 \ifx #1\expandafter \@firstoftwo
 \else \expandafter \@secondoftwo
 \fi
}%
\providecommand \natexlab [1]{#1}%
\providecommand \enquote  [1]{``#1''}%
\providecommand \bibnamefont  [1]{#1}%
\providecommand \bibfnamefont [1]{#1}%
\providecommand \citenamefont [1]{#1}%
\providecommand \href@noop [0]{\@secondoftwo}%
\providecommand \href [0]{\begingroup \@sanitize@url \@href}%
\providecommand \@href[1]{\@@startlink{#1}\@@href}%
\providecommand \@@href[1]{\endgroup#1\@@endlink}%
\providecommand \@sanitize@url [0]{\catcode `\\12\catcode `\$12\catcode
  `\&12\catcode `\#12\catcode `\^12\catcode `\_12\catcode `\%12\relax}%
\providecommand \@@startlink[1]{}%
\providecommand \@@endlink[0]{}%
\providecommand \url  [0]{\begingroup\@sanitize@url \@url }%
\providecommand \@url [1]{\endgroup\@href {#1}{\urlprefix }}%
\providecommand \urlprefix  [0]{URL }%
\providecommand \Eprint [0]{\href }%
\providecommand \doibase [0]{https://doi.org/}%
\providecommand \selectlanguage [0]{\@gobble}%
\providecommand \bibinfo  [0]{\@secondoftwo}%
\providecommand \bibfield  [0]{\@secondoftwo}%
\providecommand \translation [1]{[#1]}%
\providecommand \BibitemOpen [0]{}%
\providecommand \bibitemStop [0]{}%
\providecommand \bibitemNoStop [0]{.\EOS\space}%
\providecommand \EOS [0]{\spacefactor3000\relax}%
\providecommand \BibitemShut  [1]{\csname bibitem#1\endcsname}%
\let\auto@bib@innerbib\@empty
\bibitem [{\citenamefont {Landau}(1937)}]{Landau1937}%
  \BibitemOpen
  \bibfield  {author} {\bibinfo {author} {\bibfnamefont {L.~D.}\ \bibnamefont
  {Landau}},\ }\bibfield  {title} {\bibinfo {title} {{On the theory of phase
  transitions}},\ }\href {https://doi.org/10.1016/B978-0-08-010586-4.50034-1}
  {\bibfield  {journal} {\bibinfo  {journal} {Zh. Eksp. Teor. Fiz.}\ }\textbf
  {\bibinfo {volume} {7}},\ \bibinfo {pages} {19} (\bibinfo {year}
  {1937})}\BibitemShut {NoStop}%
\bibitem [{\citenamefont {Coleman}(1985)}]{Coleman1985}%
  \BibitemOpen
  \bibfield  {author} {\bibinfo {author} {\bibfnamefont {S.}~\bibnamefont
  {Coleman}},\ }\href {https://doi.org/10.1017/CBO9780511565045} {\emph
  {\bibinfo {title} {Aspects of Symmetry: Selected Erice Lectures}}}\ (\bibinfo
   {publisher} {Cambridge University Press},\ \bibinfo {year}
  {1985})\BibitemShut {NoStop}%
\bibitem [{\citenamefont {Kong}\ and\ \citenamefont {Zheng}(2018)}]{Kong2018}%
  \BibitemOpen
  \bibfield  {author} {\bibinfo {author} {\bibfnamefont {L.}~\bibnamefont
  {Kong}}\ and\ \bibinfo {author} {\bibfnamefont {H.}~\bibnamefont {Zheng}},\
  }\bibfield  {title} {\bibinfo {title} {Gapless edges of 2d topological orders
  and enriched monoidal categories},\ }\href
  {https://doi.org/https://doi.org/10.1016/j.nuclphysb.2017.12.007} {\bibfield
  {journal} {\bibinfo  {journal} {Nuclear Physics B}\ }\textbf {\bibinfo
  {volume} {927}},\ \bibinfo {pages} {140} (\bibinfo {year}
  {2018})}\BibitemShut {NoStop}%
\bibitem [{\citenamefont {Kong}\ and\ \citenamefont {Zheng}(2020)}]{Kong2020}%
  \BibitemOpen
  \bibfield  {author} {\bibinfo {author} {\bibfnamefont {L.}~\bibnamefont
  {Kong}}\ and\ \bibinfo {author} {\bibfnamefont {H.}~\bibnamefont {Zheng}},\
  }\bibfield  {title} {\bibinfo {title} {{A mathematical theory of gapless
  edges of 2d topological orders. Part I}},\ }\href
  {https://doi.org/10.1007/JHEP02(2020)150} {\bibfield  {journal} {\bibinfo
  {journal} {Journal of High Energy Physics}\ }\textbf {\bibinfo {volume}
  {2020}},\ \bibinfo {pages} {150} (\bibinfo {year} {2020})}\BibitemShut
  {NoStop}%
\bibitem [{\citenamefont {Kong}\ and\ \citenamefont {Zheng}(2021)}]{Kong2021}%
  \BibitemOpen
  \bibfield  {author} {\bibinfo {author} {\bibfnamefont {L.}~\bibnamefont
  {Kong}}\ and\ \bibinfo {author} {\bibfnamefont {H.}~\bibnamefont {Zheng}},\
  }\bibfield  {title} {\bibinfo {title} {{A mathematical theory of gapless
  edges of 2d topological orders. Part II}},\ }\href
  {https://doi.org/https://doi.org/10.1016/j.nuclphysb.2021.115384} {\bibfield
  {journal} {\bibinfo  {journal} {Nuclear Physics B}\ }\textbf {\bibinfo
  {volume} {966}},\ \bibinfo {pages} {115384} (\bibinfo {year}
  {2021})}\BibitemShut {NoStop}%
\bibitem [{\citenamefont {Ji}\ and\ \citenamefont {Wen}(2020)}]{Ji2020}%
  \BibitemOpen
  \bibfield  {author} {\bibinfo {author} {\bibfnamefont {W.}~\bibnamefont
  {Ji}}\ and\ \bibinfo {author} {\bibfnamefont {X.-G.}\ \bibnamefont {Wen}},\
  }\bibfield  {title} {\bibinfo {title} {Categorical symmetry and noninvertible
  anomaly in symmetry-breaking and topological phase transitions},\ }\href
  {https://doi.org/10.1103/PhysRevResearch.2.033417} {\bibfield  {journal}
  {\bibinfo  {journal} {Phys. Rev. Res.}\ }\textbf {\bibinfo {volume} {2}},\
  \bibinfo {pages} {033417} (\bibinfo {year} {2020})}\BibitemShut {NoStop}%
\bibitem [{\citenamefont {Chatterjee}\ and\ \citenamefont
  {Wen}(2023)}]{Chatterjee2023}%
  \BibitemOpen
  \bibfield  {author} {\bibinfo {author} {\bibfnamefont {A.}~\bibnamefont
  {Chatterjee}}\ and\ \bibinfo {author} {\bibfnamefont {X.-G.}\ \bibnamefont
  {Wen}},\ }\bibfield  {title} {\bibinfo {title} {Symmetry as a shadow of
  topological order and a derivation of topological holographic principle},\
  }\href {https://doi.org/10.1103/PhysRevB.107.155136} {\bibfield  {journal}
  {\bibinfo  {journal} {Phys. Rev. B}\ }\textbf {\bibinfo {volume} {107}},\
  \bibinfo {pages} {155136} (\bibinfo {year} {2023})}\BibitemShut {NoStop}%
\bibitem [{\citenamefont {Ono}\ \emph {et~al.}(2003{\natexlab{a}})\citenamefont
  {Ono}, \citenamefont {Tanaka}, \citenamefont {Katori}, \citenamefont
  {Ishikawa}, \citenamefont {Mitamura},\ and\ \citenamefont {Goto}}]{Ono2003}%
  \BibitemOpen
  \bibfield  {author} {\bibinfo {author} {\bibfnamefont {T.}~\bibnamefont
  {Ono}}, \bibinfo {author} {\bibfnamefont {H.}~\bibnamefont {Tanaka}},
  \bibinfo {author} {\bibfnamefont {H.~A.}\ \bibnamefont {Katori}}, \bibinfo
  {author} {\bibfnamefont {F.}~\bibnamefont {Ishikawa}}, \bibinfo {author}
  {\bibfnamefont {H.}~\bibnamefont {Mitamura}},\ and\ \bibinfo {author}
  {\bibfnamefont {T.}~\bibnamefont {Goto}},\ }\bibfield  {title} {\bibinfo
  {title} {{Magnetization plateau in the frustrated quantum spin system
  Cs$_2$CuBr$_4$}},\ }\href@noop {} {\bibfield  {journal} {\bibinfo  {journal}
  {Phys. Rev. B}\ }\textbf {\bibinfo {volume} {67}},\ \bibinfo {pages} {104431}
  (\bibinfo {year} {2003}{\natexlab{a}})}\BibitemShut {NoStop}%
\bibitem [{\citenamefont {Samulon}\ \emph {et~al.}(2009)\citenamefont
  {Samulon}, \citenamefont {Kohama}, \citenamefont {McDonald}, \citenamefont
  {Shapiro}, \citenamefont {Al-Hassanieh}, \citenamefont {Batista},
  \citenamefont {Jaime},\ and\ \citenamefont {Fisher}}]{Samulon2009}%
  \BibitemOpen
  \bibfield  {author} {\bibinfo {author} {\bibfnamefont {E.~C.}\ \bibnamefont
  {Samulon}}, \bibinfo {author} {\bibfnamefont {Y.}~\bibnamefont {Kohama}},
  \bibinfo {author} {\bibfnamefont {R.~D.}\ \bibnamefont {McDonald}}, \bibinfo
  {author} {\bibfnamefont {M.~C.}\ \bibnamefont {Shapiro}}, \bibinfo {author}
  {\bibfnamefont {K.~A.}\ \bibnamefont {Al-Hassanieh}}, \bibinfo {author}
  {\bibfnamefont {C.~D.}\ \bibnamefont {Batista}}, \bibinfo {author}
  {\bibfnamefont {M.}~\bibnamefont {Jaime}},\ and\ \bibinfo {author}
  {\bibfnamefont {I.~R.}\ \bibnamefont {Fisher}},\ }\bibfield  {title}
  {\bibinfo {title} {{Asymmetric Quintuplet Condensation in the Frustrated
  $S=1$ Spin Dimer Compound Ba$_3$Mn$_2$O$_8$}},\ }\href
  {https://doi.org/10.1103/PhysRevLett.103.047202} {\bibfield  {journal}
  {\bibinfo  {journal} {Phys. Rev. Lett.}\ }\textbf {\bibinfo {volume} {103}},\
  \bibinfo {pages} {047202} (\bibinfo {year} {2009})}\BibitemShut {NoStop}%
\bibitem [{\citenamefont {Samulon}\ \emph {et~al.}(2008)\citenamefont
  {Samulon}, \citenamefont {Jo}, \citenamefont {Sengupta}, \citenamefont
  {Batista}, \citenamefont {Jaime}, \citenamefont {Balicas},\ and\
  \citenamefont {Fisher}}]{Samulon2008}%
  \BibitemOpen
  \bibfield  {author} {\bibinfo {author} {\bibfnamefont {E.~C.}\ \bibnamefont
  {Samulon}}, \bibinfo {author} {\bibfnamefont {Y.~J.}\ \bibnamefont {Jo}},
  \bibinfo {author} {\bibfnamefont {P.}~\bibnamefont {Sengupta}}, \bibinfo
  {author} {\bibfnamefont {C.~D.}\ \bibnamefont {Batista}}, \bibinfo {author}
  {\bibfnamefont {M.}~\bibnamefont {Jaime}}, \bibinfo {author} {\bibfnamefont
  {L.}~\bibnamefont {Balicas}},\ and\ \bibinfo {author} {\bibfnamefont {I.~R.}\
  \bibnamefont {Fisher}},\ }\bibfield  {title} {\bibinfo {title} {{Ordered
  magnetic phases of the frustrated spin-dimer compound Ba$_3$Mn$_2$O$_8$}},\
  }\href {https://doi.org/10.1103/PhysRevB.77.214441} {\bibfield  {journal}
  {\bibinfo  {journal} {Phys. Rev. B}\ }\textbf {\bibinfo {volume} {77}},\
  \bibinfo {pages} {214441} (\bibinfo {year} {2008})}\BibitemShut {NoStop}%
\bibitem [{\citenamefont {Samulon}\ \emph {et~al.}(2010)\citenamefont
  {Samulon}, \citenamefont {Al-Hassanieh}, \citenamefont {Jo}, \citenamefont
  {Shapiro}, \citenamefont {Balicas}, \citenamefont {Batista},\ and\
  \citenamefont {Fisher}}]{Samulon2010}%
  \BibitemOpen
  \bibfield  {author} {\bibinfo {author} {\bibfnamefont {E.~C.}\ \bibnamefont
  {Samulon}}, \bibinfo {author} {\bibfnamefont {K.~A.}\ \bibnamefont
  {Al-Hassanieh}}, \bibinfo {author} {\bibfnamefont {Y.~J.}\ \bibnamefont
  {Jo}}, \bibinfo {author} {\bibfnamefont {M.~C.}\ \bibnamefont {Shapiro}},
  \bibinfo {author} {\bibfnamefont {L.}~\bibnamefont {Balicas}}, \bibinfo
  {author} {\bibfnamefont {C.~D.}\ \bibnamefont {Batista}},\ and\ \bibinfo
  {author} {\bibfnamefont {I.~R.}\ \bibnamefont {Fisher}},\ }\bibfield  {title}
  {\bibinfo {title} {{Anisotropic phase diagram of the frustrated spin dimer
  compound Ba$_3$Mn$_2$O$_8$}},\ }\href
  {https://doi.org/10.1103/PhysRevB.81.104421} {\bibfield  {journal} {\bibinfo
  {journal} {Phys. Rev. B}\ }\textbf {\bibinfo {volume} {81}},\ \bibinfo
  {pages} {104421} (\bibinfo {year} {2010})}\BibitemShut {NoStop}%
\bibitem [{\citenamefont {Shirata}\ \emph {et~al.}(2012)\citenamefont
  {Shirata}, \citenamefont {Tanaka}, \citenamefont {Matsuo},\ and\
  \citenamefont {Kindo}}]{Shirata2012}%
  \BibitemOpen
  \bibfield  {author} {\bibinfo {author} {\bibfnamefont {Y.}~\bibnamefont
  {Shirata}}, \bibinfo {author} {\bibfnamefont {H.}~\bibnamefont {Tanaka}},
  \bibinfo {author} {\bibfnamefont {A.}~\bibnamefont {Matsuo}},\ and\ \bibinfo
  {author} {\bibfnamefont {K.}~\bibnamefont {Kindo}},\ }\bibfield  {title}
  {\bibinfo {title} {{Experimental realization of a spin-1/2 triangular-lattice
  Heisenberg antiferromagnet}},\ }\href@noop {} {\bibfield  {journal} {\bibinfo
   {journal} {Phys. Rev. Lett.}\ }\textbf {\bibinfo {volume} {108}},\ \bibinfo
  {pages} {057205} (\bibinfo {year} {2012})}\BibitemShut {NoStop}%
\bibitem [{\citenamefont {Bordelon}\ \emph {et~al.}(2019)\citenamefont
  {Bordelon}, \citenamefont {Kenney}, \citenamefont {Liu}, \citenamefont
  {Hogan}, \citenamefont {Posthuma}, \citenamefont {Kavand}, \citenamefont
  {Lyu}, \citenamefont {Sherwin}, \citenamefont {Butch}, \citenamefont {Brown}
  \emph {et~al.}}]{Bordelon2019}%
  \BibitemOpen
  \bibfield  {author} {\bibinfo {author} {\bibfnamefont {M.~M.}\ \bibnamefont
  {Bordelon}}, \bibinfo {author} {\bibfnamefont {E.}~\bibnamefont {Kenney}},
  \bibinfo {author} {\bibfnamefont {C.}~\bibnamefont {Liu}}, \bibinfo {author}
  {\bibfnamefont {T.}~\bibnamefont {Hogan}}, \bibinfo {author} {\bibfnamefont
  {L.}~\bibnamefont {Posthuma}}, \bibinfo {author} {\bibfnamefont
  {M.}~\bibnamefont {Kavand}}, \bibinfo {author} {\bibfnamefont
  {Y.}~\bibnamefont {Lyu}}, \bibinfo {author} {\bibfnamefont {M.}~\bibnamefont
  {Sherwin}}, \bibinfo {author} {\bibfnamefont {N.~P.}\ \bibnamefont {Butch}},
  \bibinfo {author} {\bibfnamefont {C.}~\bibnamefont {Brown}}, \emph {et~al.},\
  }\bibfield  {title} {\bibinfo {title} {{Field-tunable quantum disordered
  ground state in the triangular-lattice antiferromagnet NaYbO$_2$}},\
  }\href@noop {} {\bibfield  {journal} {\bibinfo  {journal} {Nature Physics}\
  }\textbf {\bibinfo {volume} {15}},\ \bibinfo {pages} {1058} (\bibinfo {year}
  {2019})}\BibitemShut {NoStop}%
\bibitem [{\citenamefont {Sheng}\ \emph {et~al.}(2022)\citenamefont {Sheng},
  \citenamefont {Wang}, \citenamefont {Candini}, \citenamefont {Jiang},
  \citenamefont {Huang}, \citenamefont {Xi}, \citenamefont {Zhao},
  \citenamefont {Ge}, \citenamefont {Zhao}, \citenamefont {Fu}, \citenamefont
  {Ren}, \citenamefont {Yang}, \citenamefont {Miao}, \citenamefont {Tong},
  \citenamefont {Yu}, \citenamefont {Wang}, \citenamefont {Liu}, \citenamefont
  {Kofu}, \citenamefont {Mole}, \citenamefont {Biasiol}, \citenamefont {Yu},
  \citenamefont {Zaliznyak}, \citenamefont {Mei},\ and\ \citenamefont
  {Wu}}]{Sheng2022}%
  \BibitemOpen
  \bibfield  {author} {\bibinfo {author} {\bibfnamefont {J.}~\bibnamefont
  {Sheng}}, \bibinfo {author} {\bibfnamefont {L.}~\bibnamefont {Wang}},
  \bibinfo {author} {\bibfnamefont {A.}~\bibnamefont {Candini}}, \bibinfo
  {author} {\bibfnamefont {W.}~\bibnamefont {Jiang}}, \bibinfo {author}
  {\bibfnamefont {L.}~\bibnamefont {Huang}}, \bibinfo {author} {\bibfnamefont
  {B.}~\bibnamefont {Xi}}, \bibinfo {author} {\bibfnamefont {J.}~\bibnamefont
  {Zhao}}, \bibinfo {author} {\bibfnamefont {H.}~\bibnamefont {Ge}}, \bibinfo
  {author} {\bibfnamefont {N.}~\bibnamefont {Zhao}}, \bibinfo {author}
  {\bibfnamefont {Y.}~\bibnamefont {Fu}}, \bibinfo {author} {\bibfnamefont
  {J.}~\bibnamefont {Ren}}, \bibinfo {author} {\bibfnamefont {J.}~\bibnamefont
  {Yang}}, \bibinfo {author} {\bibfnamefont {P.}~\bibnamefont {Miao}}, \bibinfo
  {author} {\bibfnamefont {X.}~\bibnamefont {Tong}}, \bibinfo {author}
  {\bibfnamefont {D.}~\bibnamefont {Yu}}, \bibinfo {author} {\bibfnamefont
  {S.}~\bibnamefont {Wang}}, \bibinfo {author} {\bibfnamefont {Q.}~\bibnamefont
  {Liu}}, \bibinfo {author} {\bibfnamefont {M.}~\bibnamefont {Kofu}}, \bibinfo
  {author} {\bibfnamefont {R.}~\bibnamefont {Mole}}, \bibinfo {author}
  {\bibfnamefont {G.}~\bibnamefont {Biasiol}}, \bibinfo {author} {\bibfnamefont
  {D.}~\bibnamefont {Yu}}, \bibinfo {author} {\bibfnamefont {I.~A.}\
  \bibnamefont {Zaliznyak}}, \bibinfo {author} {\bibfnamefont {J.-W.}\
  \bibnamefont {Mei}},\ and\ \bibinfo {author} {\bibfnamefont {L.}~\bibnamefont
  {Wu}},\ }\bibfield  {title} {\bibinfo {title} {{Two-dimensional quantum
  universality in the spin-1/2 triangular-lattice quantum antiferromagnet
  Na$_2$BaCo(PO$_4$)$_2$}},\ }\href {https://doi.org/10.1073/pnas.2211193119}
  {\bibfield  {journal} {\bibinfo  {journal} {Proceedings of the National
  Academy of Sciences}\ }\textbf {\bibinfo {volume} {119}},\ \bibinfo {pages}
  {e2211193119} (\bibinfo {year} {2022})}\BibitemShut {NoStop}%
\bibitem [{\citenamefont {Sheng}\ \emph {et~al.}(2023)\citenamefont {Sheng},
  \citenamefont {Mei}, \citenamefont {Wang}, \citenamefont {Jiang},
  \citenamefont {Xu}, \citenamefont {Ge}, \citenamefont {Zhao}, \citenamefont
  {Li}, \citenamefont {Candini}, \citenamefont {Xi}, \citenamefont {Zhao},
  \citenamefont {Fu}, \citenamefont {Yang}, \citenamefont {Zhang},
  \citenamefont {Biasiol}, \citenamefont {Wang}, \citenamefont {Zhu},
  \citenamefont {Miao}, \citenamefont {Tong}, \citenamefont {Yu}, \citenamefont
  {Mole}, \citenamefont {Ma}, \citenamefont {Zhang}, \citenamefont {Ouyang},
  \citenamefont {Tong}, \citenamefont {Podlesnyak}, \citenamefont {Wang},
  \citenamefont {Ye}, \citenamefont {Yu}, \citenamefont {Wu},\ and\
  \citenamefont {Wang}}]{Sheng2023}%
  \BibitemOpen
  \bibfield  {author} {\bibinfo {author} {\bibfnamefont {J.}~\bibnamefont
  {Sheng}}, \bibinfo {author} {\bibfnamefont {J.-W.}\ \bibnamefont {Mei}},
  \bibinfo {author} {\bibfnamefont {L.}~\bibnamefont {Wang}}, \bibinfo {author}
  {\bibfnamefont {W.}~\bibnamefont {Jiang}}, \bibinfo {author} {\bibfnamefont
  {L.}~\bibnamefont {Xu}}, \bibinfo {author} {\bibfnamefont {H.}~\bibnamefont
  {Ge}}, \bibinfo {author} {\bibfnamefont {N.}~\bibnamefont {Zhao}}, \bibinfo
  {author} {\bibfnamefont {T.}~\bibnamefont {Li}}, \bibinfo {author}
  {\bibfnamefont {A.}~\bibnamefont {Candini}}, \bibinfo {author} {\bibfnamefont
  {B.}~\bibnamefont {Xi}}, \bibinfo {author} {\bibfnamefont {J.}~\bibnamefont
  {Zhao}}, \bibinfo {author} {\bibfnamefont {Y.}~\bibnamefont {Fu}}, \bibinfo
  {author} {\bibfnamefont {J.}~\bibnamefont {Yang}}, \bibinfo {author}
  {\bibfnamefont {Y.}~\bibnamefont {Zhang}}, \bibinfo {author} {\bibfnamefont
  {G.}~\bibnamefont {Biasiol}}, \bibinfo {author} {\bibfnamefont
  {S.}~\bibnamefont {Wang}}, \bibinfo {author} {\bibfnamefont {J.}~\bibnamefont
  {Zhu}}, \bibinfo {author} {\bibfnamefont {P.}~\bibnamefont {Miao}}, \bibinfo
  {author} {\bibfnamefont {X.}~\bibnamefont {Tong}}, \bibinfo {author}
  {\bibfnamefont {D.}~\bibnamefont {Yu}}, \bibinfo {author} {\bibfnamefont
  {R.}~\bibnamefont {Mole}}, \bibinfo {author} {\bibfnamefont {L.}~\bibnamefont
  {Ma}}, \bibinfo {author} {\bibfnamefont {Z.}~\bibnamefont {Zhang}}, \bibinfo
  {author} {\bibfnamefont {Z.}~\bibnamefont {Ouyang}}, \bibinfo {author}
  {\bibfnamefont {W.}~\bibnamefont {Tong}}, \bibinfo {author} {\bibfnamefont
  {A.}~\bibnamefont {Podlesnyak}}, \bibinfo {author} {\bibfnamefont
  {L.}~\bibnamefont {Wang}}, \bibinfo {author} {\bibfnamefont {F.}~\bibnamefont
  {Ye}}, \bibinfo {author} {\bibfnamefont {D.}~\bibnamefont {Yu}}, \bibinfo
  {author} {\bibfnamefont {L.}~\bibnamefont {Wu}},\ and\ \bibinfo {author}
  {\bibfnamefont {Z.}~\bibnamefont {Wang}},\ }\href@noop {} {\bibinfo {title}
  {Bose-einstein condensation of a two-magnon bound state in a spin-one
  triangular lattice}} (\bibinfo {year} {2023}),\ \Eprint
  {https://arxiv.org/abs/2306.09695} {arXiv:2306.09695 [cond-mat.str-el]}
  \BibitemShut {NoStop}%
\bibitem [{\citenamefont {Seabra}\ \emph {et~al.}(2011)\citenamefont {Seabra},
  \citenamefont {Momoi}, \citenamefont {Sindzingre},\ and\ \citenamefont
  {Shannon}}]{Seabra2011}%
  \BibitemOpen
  \bibfield  {author} {\bibinfo {author} {\bibfnamefont {L.}~\bibnamefont
  {Seabra}}, \bibinfo {author} {\bibfnamefont {T.}~\bibnamefont {Momoi}},
  \bibinfo {author} {\bibfnamefont {P.}~\bibnamefont {Sindzingre}},\ and\
  \bibinfo {author} {\bibfnamefont {N.}~\bibnamefont {Shannon}},\ }\bibfield
  {title} {\bibinfo {title} {Phase diagram of the classical heisenberg
  antiferromagnet on a triangular lattice in an applied magnetic field},\
  }\href {https://doi.org/10.1103/PhysRevB.84.214418} {\bibfield  {journal}
  {\bibinfo  {journal} {Phys. Rev. B}\ }\textbf {\bibinfo {volume} {84}},\
  \bibinfo {pages} {214418} (\bibinfo {year} {2011})}\BibitemShut {NoStop}%
\bibitem [{\citenamefont {Yamamoto}\ \emph {et~al.}(2014)\citenamefont
  {Yamamoto}, \citenamefont {Marmorini},\ and\ \citenamefont
  {Danshita}}]{Yamamoto2014}%
  \BibitemOpen
  \bibfield  {author} {\bibinfo {author} {\bibfnamefont {D.}~\bibnamefont
  {Yamamoto}}, \bibinfo {author} {\bibfnamefont {G.}~\bibnamefont
  {Marmorini}},\ and\ \bibinfo {author} {\bibfnamefont {I.}~\bibnamefont
  {Danshita}},\ }\bibfield  {title} {\bibinfo {title} {{Quantum Phase Diagram
  of the Triangular-Lattice XXZ Model in a Magnetic Field}},\ }\href@noop {}
  {\bibfield  {journal} {\bibinfo  {journal} {Phys. Rev. Lett.}\ }\textbf
  {\bibinfo {volume} {112}},\ \bibinfo {pages} {127203} (\bibinfo {year}
  {2014})}\BibitemShut {NoStop}%
\bibitem [{\citenamefont {Yamamoto}\ \emph {et~al.}(2019)\citenamefont
  {Yamamoto}, \citenamefont {Marmorini}, \citenamefont {Tabata}, \citenamefont
  {Sakakura},\ and\ \citenamefont {Danshita}}]{Yamamoto2019}%
  \BibitemOpen
  \bibfield  {author} {\bibinfo {author} {\bibfnamefont {D.}~\bibnamefont
  {Yamamoto}}, \bibinfo {author} {\bibfnamefont {G.}~\bibnamefont {Marmorini}},
  \bibinfo {author} {\bibfnamefont {M.}~\bibnamefont {Tabata}}, \bibinfo
  {author} {\bibfnamefont {K.}~\bibnamefont {Sakakura}},\ and\ \bibinfo
  {author} {\bibfnamefont {I.}~\bibnamefont {Danshita}},\ }\bibfield  {title}
  {\bibinfo {title} {{Magnetism driven by the interplay of fluctuations and
  frustration in the easy-axis triangular XXZ model with transverse fields}},\
  }\href@noop {} {\bibfield  {journal} {\bibinfo  {journal} {Phys. Rev. B}\
  }\textbf {\bibinfo {volume} {100}},\ \bibinfo {pages} {140410} (\bibinfo
  {year} {2019})}\BibitemShut {NoStop}%
\bibitem [{\citenamefont {Gvozdikova}\ \emph {et~al.}(2011)\citenamefont
  {Gvozdikova}, \citenamefont {Melchy},\ and\ \citenamefont
  {Zhitomirsky}}]{Gvozdikova2011}%
  \BibitemOpen
  \bibfield  {author} {\bibinfo {author} {\bibfnamefont {M.~V.}\ \bibnamefont
  {Gvozdikova}}, \bibinfo {author} {\bibfnamefont {P.-E.}\ \bibnamefont
  {Melchy}},\ and\ \bibinfo {author} {\bibfnamefont {M.~E.}\ \bibnamefont
  {Zhitomirsky}},\ }\bibfield  {title} {\bibinfo {title} {Magnetic phase
  diagrams of classical triangular and kagome antiferromagnets},\ }\href
  {https://doi.org/10.1088/0953-8984/23/16/164209} {\bibfield  {journal}
  {\bibinfo  {journal} {Journal of Physics: Condensed Matter}\ }\textbf
  {\bibinfo {volume} {23}},\ \bibinfo {pages} {164209} (\bibinfo {year}
  {2011})}\BibitemShut {NoStop}%
\bibitem [{\citenamefont {Kermarrec}\ \emph {et~al.}(2021)\citenamefont
  {Kermarrec}, \citenamefont {Kumar}, \citenamefont {Bernard}, \citenamefont
  {H\'enaff}, \citenamefont {Mendels}, \citenamefont {Bert}, \citenamefont
  {Paulose}, \citenamefont {Hazra},\ and\ \citenamefont
  {Koteswararao}}]{Kermarrec2021}%
  \BibitemOpen
  \bibfield  {author} {\bibinfo {author} {\bibfnamefont {E.}~\bibnamefont
  {Kermarrec}}, \bibinfo {author} {\bibfnamefont {R.}~\bibnamefont {Kumar}},
  \bibinfo {author} {\bibfnamefont {G.}~\bibnamefont {Bernard}}, \bibinfo
  {author} {\bibfnamefont {R.}~\bibnamefont {H\'enaff}}, \bibinfo {author}
  {\bibfnamefont {P.}~\bibnamefont {Mendels}}, \bibinfo {author} {\bibfnamefont
  {F.}~\bibnamefont {Bert}}, \bibinfo {author} {\bibfnamefont {P.~L.}\
  \bibnamefont {Paulose}}, \bibinfo {author} {\bibfnamefont {B.~K.}\
  \bibnamefont {Hazra}},\ and\ \bibinfo {author} {\bibfnamefont
  {B.}~\bibnamefont {Koteswararao}},\ }\bibfield  {title} {\bibinfo {title}
  {{Classical Spin Liquid State in the $S=\frac{5}{2}$ Heisenberg Kagome
  Antiferromagnet Li$_9$Fe$_3$(P$_2$O$_7$)$_3$(PO$_4$)$_2$}},\ }\href
  {https://doi.org/10.1103/PhysRevLett.127.157202} {\bibfield  {journal}
  {\bibinfo  {journal} {Phys. Rev. Lett.}\ }\textbf {\bibinfo {volume} {127}},\
  \bibinfo {pages} {157202} (\bibinfo {year} {2021})}\BibitemShut {NoStop}%
\bibitem [{\citenamefont {Baek}\ \emph {et~al.}(2017)\citenamefont {Baek},
  \citenamefont {Do}, \citenamefont {Choi}, \citenamefont {Kwon}, \citenamefont
  {Wolter}, \citenamefont {Nishimoto}, \citenamefont {van~den Brink},\ and\
  \citenamefont {B\"uchner}}]{Baek2017}%
  \BibitemOpen
  \bibfield  {author} {\bibinfo {author} {\bibfnamefont {S.-H.}\ \bibnamefont
  {Baek}}, \bibinfo {author} {\bibfnamefont {S.-H.}\ \bibnamefont {Do}},
  \bibinfo {author} {\bibfnamefont {K.-Y.}\ \bibnamefont {Choi}}, \bibinfo
  {author} {\bibfnamefont {Y.~S.}\ \bibnamefont {Kwon}}, \bibinfo {author}
  {\bibfnamefont {A.~U.~B.}\ \bibnamefont {Wolter}}, \bibinfo {author}
  {\bibfnamefont {S.}~\bibnamefont {Nishimoto}}, \bibinfo {author}
  {\bibfnamefont {J.}~\bibnamefont {van~den Brink}},\ and\ \bibinfo {author}
  {\bibfnamefont {B.}~\bibnamefont {B\"uchner}},\ }\bibfield  {title} {\bibinfo
  {title} {{Evidence for a Field-Induced Quantum Spin Liquid in
  $\ensuremath{\alpha}$-${\mathrm{RuCl}}_{3}$}},\ }\href
  {https://doi.org/10.1103/PhysRevLett.119.037201} {\bibfield  {journal}
  {\bibinfo  {journal} {Phys. Rev. Lett.}\ }\textbf {\bibinfo {volume} {119}},\
  \bibinfo {pages} {037201} (\bibinfo {year} {2017})}\BibitemShut {NoStop}%
\bibitem [{\citenamefont {Smith}\ \emph {et~al.}(2023)\citenamefont {Smith},
  \citenamefont {Dudemaine}, \citenamefont {Placke}, \citenamefont {Sch\"afer},
  \citenamefont {Yahne}, \citenamefont {DeLazzer}, \citenamefont {Fitterman},
  \citenamefont {Beare}, \citenamefont {Gaudet}, \citenamefont {Buhariwalla},
  \citenamefont {Podlesnyak}, \citenamefont {Xu}, \citenamefont {Clancy},
  \citenamefont {Movshovich}, \citenamefont {Luke}, \citenamefont {Ross},
  \citenamefont {Moessner}, \citenamefont {Benton}, \citenamefont {Bianchi},\
  and\ \citenamefont {Gaulin}}]{Smith2023}%
  \BibitemOpen
  \bibfield  {author} {\bibinfo {author} {\bibfnamefont {E.~M.}\ \bibnamefont
  {Smith}}, \bibinfo {author} {\bibfnamefont {J.}~\bibnamefont {Dudemaine}},
  \bibinfo {author} {\bibfnamefont {B.}~\bibnamefont {Placke}}, \bibinfo
  {author} {\bibfnamefont {R.}~\bibnamefont {Sch\"afer}}, \bibinfo {author}
  {\bibfnamefont {D.~R.}\ \bibnamefont {Yahne}}, \bibinfo {author}
  {\bibfnamefont {T.}~\bibnamefont {DeLazzer}}, \bibinfo {author}
  {\bibfnamefont {A.}~\bibnamefont {Fitterman}}, \bibinfo {author}
  {\bibfnamefont {J.}~\bibnamefont {Beare}}, \bibinfo {author} {\bibfnamefont
  {J.}~\bibnamefont {Gaudet}}, \bibinfo {author} {\bibfnamefont {C.~R.~C.}\
  \bibnamefont {Buhariwalla}}, \bibinfo {author} {\bibfnamefont
  {A.}~\bibnamefont {Podlesnyak}}, \bibinfo {author} {\bibfnamefont
  {G.}~\bibnamefont {Xu}}, \bibinfo {author} {\bibfnamefont {J.~P.}\
  \bibnamefont {Clancy}}, \bibinfo {author} {\bibfnamefont {R.}~\bibnamefont
  {Movshovich}}, \bibinfo {author} {\bibfnamefont {G.~M.}\ \bibnamefont
  {Luke}}, \bibinfo {author} {\bibfnamefont {K.~A.}\ \bibnamefont {Ross}},
  \bibinfo {author} {\bibfnamefont {R.}~\bibnamefont {Moessner}}, \bibinfo
  {author} {\bibfnamefont {O.}~\bibnamefont {Benton}}, \bibinfo {author}
  {\bibfnamefont {A.~D.}\ \bibnamefont {Bianchi}},\ and\ \bibinfo {author}
  {\bibfnamefont {B.~D.}\ \bibnamefont {Gaulin}},\ }\bibfield  {title}
  {\bibinfo {title} {{Quantum spin ice response to a magnetic field in the
  dipole-octupole pyrochlore
  ${\mathrm{Ce}}_{2}{\mathrm{Zr}}_{2}{\mathrm{O}}_{7}$}},\ }\href
  {https://doi.org/10.1103/PhysRevB.108.054438} {\bibfield  {journal} {\bibinfo
   {journal} {Phys. Rev. B}\ }\textbf {\bibinfo {volume} {108}},\ \bibinfo
  {pages} {054438} (\bibinfo {year} {2023})}\BibitemShut {NoStop}%
\bibitem [{\citenamefont {Cui}\ \emph {et~al.}(2023)\citenamefont {Cui},
  \citenamefont {Liu}, \citenamefont {Lin}, \citenamefont {Wu}, \citenamefont
  {Hong}, \citenamefont {Liu}, \citenamefont {Li}, \citenamefont {Hu},
  \citenamefont {Xi}, \citenamefont {Li}, \citenamefont {Yu}, \citenamefont
  {Sandvik},\ and\ \citenamefont {Yu}}]{Cui2023}%
  \BibitemOpen
  \bibfield  {author} {\bibinfo {author} {\bibfnamefont {Y.}~\bibnamefont
  {Cui}}, \bibinfo {author} {\bibfnamefont {L.}~\bibnamefont {Liu}}, \bibinfo
  {author} {\bibfnamefont {H.}~\bibnamefont {Lin}}, \bibinfo {author}
  {\bibfnamefont {K.-H.}\ \bibnamefont {Wu}}, \bibinfo {author} {\bibfnamefont
  {W.}~\bibnamefont {Hong}}, \bibinfo {author} {\bibfnamefont {X.}~\bibnamefont
  {Liu}}, \bibinfo {author} {\bibfnamefont {C.}~\bibnamefont {Li}}, \bibinfo
  {author} {\bibfnamefont {Z.}~\bibnamefont {Hu}}, \bibinfo {author}
  {\bibfnamefont {N.}~\bibnamefont {Xi}}, \bibinfo {author} {\bibfnamefont
  {S.}~\bibnamefont {Li}}, \bibinfo {author} {\bibfnamefont {R.}~\bibnamefont
  {Yu}}, \bibinfo {author} {\bibfnamefont {A.~W.}\ \bibnamefont {Sandvik}},\
  and\ \bibinfo {author} {\bibfnamefont {W.}~\bibnamefont {Yu}},\ }\bibfield
  {title} {\bibinfo {title} {{Proximate deconfined quantum critical point in
  SrCu$_2$(BO$_3$)$_2$}},\ }\href {https://doi.org/10.1126/science.adc9487}
  {\bibfield  {journal} {\bibinfo  {journal} {Science}\ }\textbf {\bibinfo
  {volume} {380}},\ \bibinfo {pages} {1179} (\bibinfo {year}
  {2023})}\BibitemShut {NoStop}%
\bibitem [{\citenamefont {Balents}(2010)}]{Balents2010}%
  \BibitemOpen
  \bibfield  {author} {\bibinfo {author} {\bibfnamefont {L.}~\bibnamefont
  {Balents}},\ }\bibfield  {title} {\bibinfo {title} {Spin liquids in
  frustrated magnets},\ }\href {https://doi.org/10.1038/nature08917} {\bibfield
   {journal} {\bibinfo  {journal} {Nature}\ }\textbf {\bibinfo {volume}
  {464}},\ \bibinfo {pages} {199} (\bibinfo {year} {2010})},\ \bibinfo {note}
  {1}\BibitemShut {NoStop}%
\bibitem [{\citenamefont {Moessner}\ and\ \citenamefont
  {Ramirez}(2006)}]{Moessner2006}%
  \BibitemOpen
  \bibfield  {author} {\bibinfo {author} {\bibfnamefont {R.}~\bibnamefont
  {Moessner}}\ and\ \bibinfo {author} {\bibfnamefont {A.~P.}\ \bibnamefont
  {Ramirez}},\ }\bibfield  {title} {\bibinfo {title} {Geometrical
  frustration},\ }\href {https://doi.org/10.1063/1.2186278} {\bibfield
  {journal} {\bibinfo  {journal} {Physics Today}\ }\textbf {\bibinfo {volume}
  {59}},\ \bibinfo {pages} {24} (\bibinfo {year} {2006})}\BibitemShut {NoStop}%
\bibitem [{\citenamefont {Zapf}\ \emph {et~al.}(2014)\citenamefont {Zapf},
  \citenamefont {Jaime},\ and\ \citenamefont {Batista}}]{Zapf2014}%
  \BibitemOpen
  \bibfield  {author} {\bibinfo {author} {\bibfnamefont {V.}~\bibnamefont
  {Zapf}}, \bibinfo {author} {\bibfnamefont {M.}~\bibnamefont {Jaime}},\ and\
  \bibinfo {author} {\bibfnamefont {C.~D.}\ \bibnamefont {Batista}},\
  }\bibfield  {title} {\bibinfo {title} {Bose-einstein condensation in quantum
  magnets},\ }\href {https://doi.org/10.1103/RevModPhys.86.563} {\bibfield
  {journal} {\bibinfo  {journal} {Rev. Mod. Phys.}\ }\textbf {\bibinfo {volume}
  {86}},\ \bibinfo {pages} {563} (\bibinfo {year} {2014})}\BibitemShut
  {NoStop}%
\bibitem [{\citenamefont {Ono}\ \emph {et~al.}(2003{\natexlab{b}})\citenamefont
  {Ono}, \citenamefont {Tanaka}, \citenamefont {Aruga~Katori}, \citenamefont
  {Ishikawa}, \citenamefont {Mitamura},\ and\ \citenamefont {Goto}}]{Ono2007}%
  \BibitemOpen
  \bibfield  {author} {\bibinfo {author} {\bibfnamefont {T.}~\bibnamefont
  {Ono}}, \bibinfo {author} {\bibfnamefont {H.}~\bibnamefont {Tanaka}},
  \bibinfo {author} {\bibfnamefont {H.}~\bibnamefont {Aruga~Katori}}, \bibinfo
  {author} {\bibfnamefont {F.}~\bibnamefont {Ishikawa}}, \bibinfo {author}
  {\bibfnamefont {H.}~\bibnamefont {Mitamura}},\ and\ \bibinfo {author}
  {\bibfnamefont {T.}~\bibnamefont {Goto}},\ }\bibfield  {title} {\bibinfo
  {title} {{Magnetization plateau in the frustrated quantum spin system
  ${\mathrm{Cs}}_{2}{\mathrm{CuBr}}_{4}$}},\ }\href
  {https://doi.org/10.1103/PhysRevB.67.104431} {\bibfield  {journal} {\bibinfo
  {journal} {Physical Review B}\ }\textbf {\bibinfo {volume} {67}},\ \bibinfo
  {pages} {104431} (\bibinfo {year} {2003}{\natexlab{b}})}\BibitemShut
  {NoStop}%
\bibitem [{\citenamefont {Alicea}\ \emph {et~al.}(2009)\citenamefont {Alicea},
  \citenamefont {Chubukov},\ and\ \citenamefont {Starykh}}]{Alicea2009}%
  \BibitemOpen
  \bibfield  {author} {\bibinfo {author} {\bibfnamefont {J.}~\bibnamefont
  {Alicea}}, \bibinfo {author} {\bibfnamefont {A.~V.}\ \bibnamefont
  {Chubukov}},\ and\ \bibinfo {author} {\bibfnamefont {O.~A.}\ \bibnamefont
  {Starykh}},\ }\bibfield  {title} {\bibinfo {title} {{Quantum Stabilization of
  the $1/3$-Magnetization Plateau in ${\mathrm{Cs}}_{2}{\mathrm{CuBr}}_{4}$}},\
  }\href {https://doi.org/10.1103/PhysRevLett.102.137201} {\bibfield  {journal}
  {\bibinfo  {journal} {Phys. Rev. Lett}\ }\textbf {\bibinfo {volume} {102}},\
  \bibinfo {pages} {137201} (\bibinfo {year} {2009})}\BibitemShut {NoStop}%
\bibitem [{\citenamefont {Jeon}\ \emph {et~al.}(2024)\citenamefont {Jeon},
  \citenamefont {Wulferding}, \citenamefont {Choi}, \citenamefont {Lee},
  \citenamefont {Nam}, \citenamefont {Kim}, \citenamefont {Lee}, \citenamefont
  {Jang}, \citenamefont {Park}, \citenamefont {Lee}, \citenamefont {Choi},
  \citenamefont {Lee}, \citenamefont {Nojiri},\ and\ \citenamefont
  {Choi}}]{Jeon2024}%
  \BibitemOpen
  \bibfield  {author} {\bibinfo {author} {\bibfnamefont {S.}~\bibnamefont
  {Jeon}}, \bibinfo {author} {\bibfnamefont {D.}~\bibnamefont {Wulferding}},
  \bibinfo {author} {\bibfnamefont {Y.}~\bibnamefont {Choi}}, \bibinfo {author}
  {\bibfnamefont {S.}~\bibnamefont {Lee}}, \bibinfo {author} {\bibfnamefont
  {K.}~\bibnamefont {Nam}}, \bibinfo {author} {\bibfnamefont {K.~H.}\
  \bibnamefont {Kim}}, \bibinfo {author} {\bibfnamefont {M.}~\bibnamefont
  {Lee}}, \bibinfo {author} {\bibfnamefont {T.-H.}\ \bibnamefont {Jang}},
  \bibinfo {author} {\bibfnamefont {J.-H.}\ \bibnamefont {Park}}, \bibinfo
  {author} {\bibfnamefont {S.}~\bibnamefont {Lee}}, \bibinfo {author}
  {\bibfnamefont {S.}~\bibnamefont {Choi}}, \bibinfo {author} {\bibfnamefont
  {C.}~\bibnamefont {Lee}}, \bibinfo {author} {\bibfnamefont {H.}~\bibnamefont
  {Nojiri}},\ and\ \bibinfo {author} {\bibfnamefont {K.-Y.}\ \bibnamefont
  {Choi}},\ }\bibfield  {title} {\bibinfo {title} {One-ninth magnetization
  plateau stabilized by spin entanglement in a kagome antiferromagnet},\ }\href
  {https://doi.org/10.1038/s41567-023-02318-7} {\bibfield  {journal} {\bibinfo
  {journal} {Nature Physics}\ }\textbf {\bibinfo {volume} {20}},\ \bibinfo
  {pages} {435} (\bibinfo {year} {2024})}\BibitemShut {NoStop}%
\bibitem [{\citenamefont {Stone}\ \emph {et~al.}(2008)\citenamefont {Stone},
  \citenamefont {Lumsden}, \citenamefont {Qiu}, \citenamefont {Samulon},
  \citenamefont {Batista},\ and\ \citenamefont {Fisher}}]{Stone2008}%
  \BibitemOpen
  \bibfield  {author} {\bibinfo {author} {\bibfnamefont {M.~B.}\ \bibnamefont
  {Stone}}, \bibinfo {author} {\bibfnamefont {M.~D.}\ \bibnamefont {Lumsden}},
  \bibinfo {author} {\bibfnamefont {Y.}~\bibnamefont {Qiu}}, \bibinfo {author}
  {\bibfnamefont {E.~C.}\ \bibnamefont {Samulon}}, \bibinfo {author}
  {\bibfnamefont {C.~D.}\ \bibnamefont {Batista}},\ and\ \bibinfo {author}
  {\bibfnamefont {I.~R.}\ \bibnamefont {Fisher}},\ }\bibfield  {title}
  {\bibinfo {title} {{Dispersive magnetic excitations in the $S=1$
  antiferromagnet Ba$_3$Mn$_2$O$_8$}},\ }\href
  {https://doi.org/10.1103/PhysRevB.77.134406} {\bibfield  {journal} {\bibinfo
  {journal} {Phys. Rev. B}\ }\textbf {\bibinfo {volume} {77}},\ \bibinfo
  {pages} {134406} (\bibinfo {year} {2008})}\BibitemShut {NoStop}%
\bibitem [{\citenamefont {Tsujii}\ \emph {et~al.}(2005)\citenamefont {Tsujii},
  \citenamefont {Andraka}, \citenamefont {Uchida}, \citenamefont {Tanaka},\
  and\ \citenamefont {Takano}}]{TsuJii2005}%
  \BibitemOpen
  \bibfield  {author} {\bibinfo {author} {\bibfnamefont {H.}~\bibnamefont
  {Tsujii}}, \bibinfo {author} {\bibfnamefont {B.}~\bibnamefont {Andraka}},
  \bibinfo {author} {\bibfnamefont {M.}~\bibnamefont {Uchida}}, \bibinfo
  {author} {\bibfnamefont {H.}~\bibnamefont {Tanaka}},\ and\ \bibinfo {author}
  {\bibfnamefont {Y.}~\bibnamefont {Takano}},\ }\bibfield  {title} {\bibinfo
  {title} {{Specific heat of the $S=1$ spin-dimer antiferromagnet
  Ba$_3$Mn$_2$O$_8$ in high magnetic fields}},\ }\href
  {https://doi.org/10.1103/PhysRevB.72.214434} {\bibfield  {journal} {\bibinfo
  {journal} {Phys. Rev. B}\ }\textbf {\bibinfo {volume} {72}},\ \bibinfo
  {pages} {214434} (\bibinfo {year} {2005})}\BibitemShut {NoStop}%
\bibitem [{\citenamefont {Wildner}(1994)}]{Wildner1994}%
  \BibitemOpen
  \bibfield  {author} {\bibinfo {author} {\bibfnamefont {M.}~\bibnamefont
  {Wildner}},\ }\bibfield  {title} {\bibinfo {title} {{Structure of
  K$_2$Co$_2$(SeO$_3$)$_3$}},\ }\href@noop {} {\bibfield  {journal} {\bibinfo
  {journal} {Acta Crystallographica Section C-crystal Structure
  Communications}\ }\textbf {\bibinfo {volume} {50}},\ \bibinfo {pages} {336}
  (\bibinfo {year} {1994})}\BibitemShut {NoStop}%
\bibitem [{\citenamefont {Zhong}\ \emph {et~al.}(2020)\citenamefont {Zhong},
  \citenamefont {Guo}, \citenamefont {Nguyen},\ and\ \citenamefont
  {Cava}}]{Zhong2020}%
  \BibitemOpen
  \bibfield  {author} {\bibinfo {author} {\bibfnamefont {R.}~\bibnamefont
  {Zhong}}, \bibinfo {author} {\bibfnamefont {S.}~\bibnamefont {Guo}}, \bibinfo
  {author} {\bibfnamefont {L.~T.}\ \bibnamefont {Nguyen}},\ and\ \bibinfo
  {author} {\bibfnamefont {R.~J.}\ \bibnamefont {Cava}},\ }\bibfield  {title}
  {\bibinfo {title} {{Frustrated spin-1/2 dimer compound
  K$_2$Co$_2$(SeO$_3$)$_3$ with easy-axis anisotropy}},\ }\href
  {https://doi.org/10.1103/PhysRevB.102.224430} {\bibfield  {journal} {\bibinfo
   {journal} {Phys. Rev. B}\ }\textbf {\bibinfo {volume} {102}},\ \bibinfo
  {pages} {224430} (\bibinfo {year} {2020})}\BibitemShut {NoStop}%
\bibitem [{\citenamefont {Sheldrick}(2015)}]{refinement}%
  \BibitemOpen
  \bibfield  {author} {\bibinfo {author} {\bibfnamefont {G.~M.}\ \bibnamefont
  {Sheldrick}},\ }\bibfield  {title} {\bibinfo {title} {Crystal structure
  refinement with shelxl},\ }\href@noop {} {\bibfield  {journal} {\bibinfo
  {journal} {Acta Crystallographica. Section C, Structural Chemistry}\ }\textbf
  {\bibinfo {volume} {71}},\ \bibinfo {pages} {3 } (\bibinfo {year}
  {2015})}\BibitemShut {NoStop}%
\bibitem [{\citenamefont {Bourhis}\ \emph {et~al.}(2015)\citenamefont
  {Bourhis}, \citenamefont {Dolomanov}, \citenamefont {Gildea}, \citenamefont
  {Howard},\ and\ \citenamefont {Puschmann}}]{refinement2}%
  \BibitemOpen
  \bibfield  {author} {\bibinfo {author} {\bibfnamefont {L.~J.}\ \bibnamefont
  {Bourhis}}, \bibinfo {author} {\bibfnamefont {O.~V.}\ \bibnamefont
  {Dolomanov}}, \bibinfo {author} {\bibfnamefont {R.~J.}\ \bibnamefont
  {Gildea}}, \bibinfo {author} {\bibfnamefont {J.~A.~K.}\ \bibnamefont
  {Howard}},\ and\ \bibinfo {author} {\bibfnamefont {H.}~\bibnamefont
  {Puschmann}},\ }\bibfield  {title} {\bibinfo {title} {The anatomy of a
  comprehensive constrained, restrained refinement program for the modern
  computing environment – olex2 dissected},\ }\href@noop {} {\bibfield
  {journal} {\bibinfo  {journal} {Acta Crystallographica. Section A,
  Foundations and Advances}\ }\textbf {\bibinfo {volume} {71}},\ \bibinfo
  {pages} {59 } (\bibinfo {year} {2015})}\BibitemShut {NoStop}%
\bibitem [{\citenamefont {Dolomanov}\ \emph {et~al.}(2009)\citenamefont
  {Dolomanov}, \citenamefont {Bourhis}, \citenamefont {Gildea}, \citenamefont
  {Howard},\ and\ \citenamefont {Puschmann}}]{Dolomanov2009}%
  \BibitemOpen
  \bibfield  {author} {\bibinfo {author} {\bibfnamefont {O.~V.}\ \bibnamefont
  {Dolomanov}}, \bibinfo {author} {\bibfnamefont {L.~J.}\ \bibnamefont
  {Bourhis}}, \bibinfo {author} {\bibfnamefont {R.~J.}\ \bibnamefont {Gildea}},
  \bibinfo {author} {\bibfnamefont {J.~A.~K.}\ \bibnamefont {Howard}},\ and\
  \bibinfo {author} {\bibfnamefont {H.}~\bibnamefont {Puschmann}},\ }\bibfield
  {title} {\bibinfo {title} {{{\it OLEX2}: a complete structure solution,
  refinement and analysis program}},\ }\href
  {https://doi.org/10.1107/S0021889808042726} {\bibfield  {journal} {\bibinfo
  {journal} {Journal of Applied Crystallography}\ }\textbf {\bibinfo {volume}
  {42}},\ \bibinfo {pages} {339} (\bibinfo {year} {2009})}\BibitemShut
  {NoStop}%
\bibitem [{\citenamefont {Kresse}\ and\ \citenamefont
  {Furthmüller}(1996)}]{Kresse1996}%
  \BibitemOpen
  \bibfield  {author} {\bibinfo {author} {\bibfnamefont {G.}~\bibnamefont
  {Kresse}}\ and\ \bibinfo {author} {\bibfnamefont {J.}~\bibnamefont
  {Furthmüller}},\ }\bibfield  {title} {\bibinfo {title} {Efficient iterative
  schemes for ab initio total-energy calculations using a plane-wave basis
  set},\ }\href {https://doi.org/10.1103/PhysRevB.54.11169} {\bibfield
  {journal} {\bibinfo  {journal} {Phys. Rev. B}\ }\textbf {\bibinfo {volume}
  {54}},\ \bibinfo {pages} {11169} (\bibinfo {year} {1996})}\BibitemShut
  {NoStop}%
\bibitem [{\citenamefont {Blöchl}(1994)}]{Blochi1994}%
  \BibitemOpen
  \bibfield  {author} {\bibinfo {author} {\bibfnamefont {P.~E.}\ \bibnamefont
  {Blöchl}},\ }\bibfield  {title} {\bibinfo {title} {Projector augmented-wave
  method},\ }\href {https://doi.org/10.1103/PhysRevB.50.17953} {\bibfield
  {journal} {\bibinfo  {journal} {Phys. Rev. B}\ }\textbf {\bibinfo {volume}
  {50}},\ \bibinfo {pages} {17953} (\bibinfo {year} {1994})}\BibitemShut
  {NoStop}%
\bibitem [{\citenamefont {Perdew}\ \emph {et~al.}(1996)\citenamefont {Perdew},
  \citenamefont {Burke},\ and\ \citenamefont {Ernzerhof}}]{Perdew1996}%
  \BibitemOpen
  \bibfield  {author} {\bibinfo {author} {\bibfnamefont {J.~P.}\ \bibnamefont
  {Perdew}}, \bibinfo {author} {\bibfnamefont {K.}~\bibnamefont {Burke}},\ and\
  \bibinfo {author} {\bibfnamefont {M.}~\bibnamefont {Ernzerhof}},\ }\bibfield
  {title} {\bibinfo {title} {Generalized gradient approximation made simple},\
  }\href {https://doi.org/10.1103/PhysRevLett.77.3865} {\bibfield  {journal}
  {\bibinfo  {journal} {Phys. Rev. Lett.}\ }\textbf {\bibinfo {volume} {77}},\
  \bibinfo {pages} {3865} (\bibinfo {year} {1996})}\BibitemShut {NoStop}%
\bibitem [{\citenamefont {Perdew}\ \emph {et~al.}(2008)\citenamefont {Perdew},
  \citenamefont {Ruzsinszky}, \citenamefont {Csonka}, \citenamefont {Vydrov},
  \citenamefont {Scuseria}, \citenamefont {Constantin}, \citenamefont {Zhou},\
  and\ \citenamefont {Burke}}]{Perdew2008}%
  \BibitemOpen
  \bibfield  {author} {\bibinfo {author} {\bibfnamefont {J.~P.}\ \bibnamefont
  {Perdew}}, \bibinfo {author} {\bibfnamefont {A.}~\bibnamefont {Ruzsinszky}},
  \bibinfo {author} {\bibfnamefont {G.~I.}\ \bibnamefont {Csonka}}, \bibinfo
  {author} {\bibfnamefont {O.~A.}\ \bibnamefont {Vydrov}}, \bibinfo {author}
  {\bibfnamefont {G.~E.}\ \bibnamefont {Scuseria}}, \bibinfo {author}
  {\bibfnamefont {L.~A.}\ \bibnamefont {Constantin}}, \bibinfo {author}
  {\bibfnamefont {X.}~\bibnamefont {Zhou}},\ and\ \bibinfo {author}
  {\bibfnamefont {K.}~\bibnamefont {Burke}},\ }\bibfield  {title} {\bibinfo
  {title} {Restoring the density-gradient expansion for exchange in solids and
  surfaces},\ }\href {https://doi.org/10.1103/PhysRevLett.100.136406}
  {\bibfield  {journal} {\bibinfo  {journal} {Phys. Rev. Lett.}\ }\textbf
  {\bibinfo {volume} {100}},\ \bibinfo {pages} {136406} (\bibinfo {year}
  {2008})}\BibitemShut {NoStop}%
\bibitem [{\citenamefont {Dudarev}\ \emph {et~al.}(1998)\citenamefont
  {Dudarev}, \citenamefont {Botton}, \citenamefont {Savrasov}, \citenamefont
  {Humphreys},\ and\ \citenamefont {Sutton}}]{Dudarev1998}%
  \BibitemOpen
  \bibfield  {author} {\bibinfo {author} {\bibfnamefont {S.~L.}\ \bibnamefont
  {Dudarev}}, \bibinfo {author} {\bibfnamefont {G.~A.}\ \bibnamefont {Botton}},
  \bibinfo {author} {\bibfnamefont {S.~Y.}\ \bibnamefont {Savrasov}}, \bibinfo
  {author} {\bibfnamefont {C.~J.}\ \bibnamefont {Humphreys}},\ and\ \bibinfo
  {author} {\bibfnamefont {A.~P.}\ \bibnamefont {Sutton}},\ }\bibfield  {title}
  {\bibinfo {title} {Electron-energy-loss spectra and the structural stability
  of nickel oxide: An lsda+u study},\ }\href
  {https://doi.org/10.1103/PhysRevB.57.1505} {\bibfield  {journal} {\bibinfo
  {journal} {Phys. Rev. B}\ }\textbf {\bibinfo {volume} {57}},\ \bibinfo
  {pages} {1505} (\bibinfo {year} {1998})}\BibitemShut {NoStop}%
\bibitem [{\citenamefont {Steiner}\ \emph {et~al.}(2016)\citenamefont
  {Steiner}, \citenamefont {Khmelevskyi}, \citenamefont {Marsmann},\ and\
  \citenamefont {Kresse}}]{Steiner2016}%
  \BibitemOpen
  \bibfield  {author} {\bibinfo {author} {\bibfnamefont {S.}~\bibnamefont
  {Steiner}}, \bibinfo {author} {\bibfnamefont {S.}~\bibnamefont
  {Khmelevskyi}}, \bibinfo {author} {\bibfnamefont {M.}~\bibnamefont
  {Marsmann}},\ and\ \bibinfo {author} {\bibfnamefont {G.}~\bibnamefont
  {Kresse}},\ }\bibfield  {title} {\bibinfo {title} {Calculation of the
  magnetic anisotropy with projected-augmented-wave methodology and the case
  study of disordered ${\mathrm{fe}}_{1\ensuremath{-}x}{\mathrm{co}}_{x}$
  alloys},\ }\href {https://doi.org/10.1103/PhysRevB.93.224425} {\bibfield
  {journal} {\bibinfo  {journal} {Phys. Rev. B}\ }\textbf {\bibinfo {volume}
  {93}},\ \bibinfo {pages} {224425} (\bibinfo {year} {2016})}\BibitemShut
  {NoStop}%
\bibitem [{\citenamefont {Goodenough}(2008)}]{Goodenough2008}%
  \BibitemOpen
  \bibfield  {author} {\bibinfo {author} {\bibfnamefont {J.~B.}\ \bibnamefont
  {Goodenough}},\ }\bibfield  {title} {\bibinfo {title}
  {{G}oodenough-{K}anamori rule},\ }\href
  {https://doi.org/10.4249/scholarpedia.7382} {\bibfield  {journal} {\bibinfo
  {journal} {Scholarpedia}\ }\textbf {\bibinfo {volume} {3}},\ \bibinfo {pages}
  {7382} (\bibinfo {year} {2008})},\ \bibinfo {note} {revision
  \#122456}\BibitemShut {NoStop}%
\bibitem [{\citenamefont {Van~Vleck}(1937)}]{VanVleck1937}%
  \BibitemOpen
  \bibfield  {author} {\bibinfo {author} {\bibfnamefont {J.~H.}\ \bibnamefont
  {Van~Vleck}},\ }\bibfield  {title} {\bibinfo {title} {{The Influence of
  Dipole‐Dipole Coupling on the Specific Heat and Susceptibility of a
  Paramagnetic Salt}},\ }\href {https://doi.org/10.1063/1.1750032} {\bibfield
  {journal} {\bibinfo  {journal} {The Journal of Chemical Physics}\ }\textbf
  {\bibinfo {volume} {5}},\ \bibinfo {pages} {320} (\bibinfo {year}
  {1937})}\BibitemShut {NoStop}%
\bibitem [{\citenamefont {Melchy}\ and\ \citenamefont
  {Zhitomirsky}(2009)}]{Melchy2009}%
  \BibitemOpen
  \bibfield  {author} {\bibinfo {author} {\bibfnamefont {P.-E.}\ \bibnamefont
  {Melchy}}\ and\ \bibinfo {author} {\bibfnamefont {M.~E.}\ \bibnamefont
  {Zhitomirsky}},\ }\bibfield  {title} {\bibinfo {title} {Interplay of
  anisotropy and frustration: Triple transitions in a triangular-lattice
  antiferromagnet},\ }\href {https://doi.org/10.1103/PhysRevB.80.064411}
  {\bibfield  {journal} {\bibinfo  {journal} {Phys. Rev. B}\ }\textbf {\bibinfo
  {volume} {80}},\ \bibinfo {pages} {064411} (\bibinfo {year}
  {2009})}\BibitemShut {NoStop}%
\bibitem [{\citenamefont {Ishii}\ \emph {et~al.}(2009)\citenamefont {Ishii},
  \citenamefont {Tanaka}, \citenamefont {Onuma}, \citenamefont {Nambu},
  \citenamefont {Tokunaga}, \citenamefont {Sakakibara}, \citenamefont
  {Kawashima}, \citenamefont {Maeno}, \citenamefont {Broholm}, \citenamefont
  {Gautreaux}, \citenamefont {Chan},\ and\ \citenamefont
  {Nakatsuji}}]{Ishii2009}%
  \BibitemOpen
  \bibfield  {author} {\bibinfo {author} {\bibfnamefont {R.}~\bibnamefont
  {Ishii}}, \bibinfo {author} {\bibfnamefont {S.}~\bibnamefont {Tanaka}},
  \bibinfo {author} {\bibfnamefont {K.}~\bibnamefont {Onuma}}, \bibinfo
  {author} {\bibfnamefont {Y.}~\bibnamefont {Nambu}}, \bibinfo {author}
  {\bibfnamefont {M.}~\bibnamefont {Tokunaga}}, \bibinfo {author}
  {\bibfnamefont {T.}~\bibnamefont {Sakakibara}}, \bibinfo {author}
  {\bibfnamefont {N.}~\bibnamefont {Kawashima}}, \bibinfo {author}
  {\bibfnamefont {Y.}~\bibnamefont {Maeno}}, \bibinfo {author} {\bibfnamefont
  {C.}~\bibnamefont {Broholm}}, \bibinfo {author} {\bibfnamefont {D.~P.}\
  \bibnamefont {Gautreaux}}, \bibinfo {author} {\bibfnamefont {J.~Y.}\
  \bibnamefont {Chan}},\ and\ \bibinfo {author} {\bibfnamefont
  {S.}~\bibnamefont {Nakatsuji}},\ }\bibfield  {title} {\bibinfo {title}
  {{Successive phase transitions and phase diagrams for the
  quasi-two-dimensional easy-axis triangular antiferromagnet
  Rb$_4$Mn(MoO$_4$)$_3$}},\ }\href@noop {} {\bibfield  {journal} {\bibinfo
  {journal} {EPL (Europhysics Letters)}\ }\textbf {\bibinfo {volume} {94}},\
  \bibinfo {pages} {17001} (\bibinfo {year} {2009})}\BibitemShut {NoStop}%
\bibitem [{\citenamefont {Lee}\ \emph {et~al.}(2014)\citenamefont {Lee},
  \citenamefont {Choi}, \citenamefont {Huang}, \citenamefont {Ma},
  \citenamefont {Dela~Cruz}, \citenamefont {Matsuda}, \citenamefont {Tian},
  \citenamefont {Dun}, \citenamefont {Dong},\ and\ \citenamefont
  {Zhou}}]{Lee2014}%
  \BibitemOpen
  \bibfield  {author} {\bibinfo {author} {\bibfnamefont {M.}~\bibnamefont
  {Lee}}, \bibinfo {author} {\bibfnamefont {E.~S.}\ \bibnamefont {Choi}},
  \bibinfo {author} {\bibfnamefont {X.}~\bibnamefont {Huang}}, \bibinfo
  {author} {\bibfnamefont {J.}~\bibnamefont {Ma}}, \bibinfo {author}
  {\bibfnamefont {C.~R.}\ \bibnamefont {Dela~Cruz}}, \bibinfo {author}
  {\bibfnamefont {M.}~\bibnamefont {Matsuda}}, \bibinfo {author} {\bibfnamefont
  {W.}~\bibnamefont {Tian}}, \bibinfo {author} {\bibfnamefont {Z.~L.}\
  \bibnamefont {Dun}}, \bibinfo {author} {\bibfnamefont {S.}~\bibnamefont
  {Dong}},\ and\ \bibinfo {author} {\bibfnamefont {H.~D.}\ \bibnamefont
  {Zhou}},\ }\bibfield  {title} {\bibinfo {title} {{Magnetic phase diagram and
  multiferroicity of Ba$_{3}$MnNb$_{2}$O$_{9}$: A spin-$\frac{5}{2}$ triangular
  lattice antiferromagnet with weak easy-axis anisotropy}},\ }\href
  {https://doi.org/10.1103/PhysRevB.90.224402} {\bibfield  {journal} {\bibinfo
  {journal} {Phys. Rev. Lett}\ }\textbf {\bibinfo {volume} {90}},\ \bibinfo
  {pages} {224402} (\bibinfo {year} {2014})}\BibitemShut {NoStop}%
\end{thebibliography}%

\end{document}